\documentclass[%
 reprint,
 amsmath,amssymb,
 aps,
showkeys]{revtex4-2}

\usepackage{graphicx}
\usepackage{dcolumn}
\usepackage{bm}
\usepackage{subfiles}
\usepackage{url}

\usepackage{breakurl}
\usepackage{hyperref}
\hypersetup{breaklinks}

\usepackage{subfigure}
\usepackage{svg}
\usepackage{enumitem}

\begin{document}

\preprint{APS/123-QED}

\title{Topological Data Analysis Detects Percolation Thresholds\\ in Arctic Melt-Pond Evolution}

\author{W. Offord, M. Coughlan, I. J. Hewitt, H. A. Harrington, G. Grindstaff*}
    \affiliation{Mathematical Institute, University of Oxford}
    \affiliation{*Department of Mathematics, University of California, Los Angeles}

\date{\today}

\begin{abstract}
During the summer melt period, ponds form on the surface of Arctic sea ice as it melts, with important consequences for ice evolution and marine ecology. Due to the \textit{ice-albedo feedback}, these {\em melt ponds} experience uneven heat absorption, and exhibit complex patterns, which has motivated the development of modelling and data analysis to understand their particular dynamics.
 We provide a multiscale shape analysis using tools from computational algebraic topology, simultaneously capturing convexity, proximity, integrity, and feature size complementing existing single-scale quantification.
 Of particular interest in modelling the ponds is a percolation threshold at which local pond structure begins merging into macroscopic features. This percolation threshold has previously been observed using fractal dimension techniques. The signed Euclidean distance transform (SEDT) is a topological encoding of heterogeneous shape in binary images, and has been previously applied to porous media for percolation as well as other material behaviours. Here we adapt the SEDT for Arctic melt pond data to give a rich characterization and computation of shape, quantifying overall melt pond development in several complementary ways, and from which classical percolation and dimension results can be extracted. 
This orientation-invariant topological approach distinguishes different dynamical network models of melt pond evolution of varying complexity.
\end{abstract}

\maketitle


\section{\label{sec:intro}Introduction}

Melt ponds on Arctic sea ice develop into complex patterns over the course of the melt season. These melt pond dynamics have been studied extensively \cite{polashenskimechanisms, coughlan2021dynamical}, as they are a key factor for overall heat absorption and melt rate of the sea ice as a whole. Melt ponds have a lower albedo than ice, and hence absorb more heat in what is known as \textit{ice-albedo feedback} \cite{SeaIceAlbedoModel, SeaIceAlbedoClimateFeedbackMechanism,shrinkingseasynthesis,nonlinearthreshold,increasingsolarheating}. Understanding such mechanisms is therefore important for predicting sea ice decline, which has been underestimated by models \cite{seaicedecline}. Moreover, melt pond patterns determine the amount of light reaching the upper ocean and hence influence the biological productivity in the upper ocean \citep{plankton,planktonfreq,lighttransmittance,spacialsolardistribution}. Understanding how these melt ponds evolve is a key step in modelling and predicting both global climate evolution and upper ocean ecology.

While broad outcome statistics like total pond coverage fraction have been studied \cite{polashenskimechanisms}, there have also been advances in understanding the finer geometry of the melt ponds themselves. \citet{FractalDimensionPonds} observes a transition in the fractal dimension of ponds as pond area increases, indicating a critical area threshold at which local pond geometry begins to combine to form more macroscopic features. This transition in fractal dimension has been reproduced in simple geometric models of melt ponds. \citet{simplerules} model the ponds by considering randomly placed circles on a plane, with ponds represented as the negative space around these circles. \citet{Ising} models melt ponds as metastable states in an Ising model.

The goal of this paper is to develop a larger range of metrics with which to analyse melt-pond imagery and model output, using tools from topological data analysis (TDA). During the past few decades, this field has developed theory and computational techniques, applying invariants from algebraic topology to robustly summarize shape and structure in complex data sets. Topological data analysis has been rapidly adopted in certain fields like biology, medicine and materials science \cite{MPHHarrington, vascular, proteins, flowestimation}, but has only recently begun to see use in geophysics \cite{LandslideDetection, weather}. 

Topology is a natural tool, bridging local and global connectivity, to study percolation behaviour such as that exhibited by melt ponds.
In \citet{RobinsSEDT}, topological techniques are developed to measure detailed information related to percolation in porous materials. \citet{flowestimation} uses these techniques to to predict flow through porous media, while \citet{statisticalinference} apply statistical techniques to the output topological descriptors in order to determine representative sample sizes; they also use machine-learning to estimate fluid flow and transport. We apply these TDA techniques to melt pond evolution, obtaining information about the geometric structure of melt ponds and percolating length scales in pond evolution models. 
This information can be used to quantify relevant geometric and topological information about the pond networks arising from models, which could lead to additional criteria for model selection beyond those already studied, such as fractal dimension, concavity, percolation and loops in pond structure, and size of ponds and channels between them. 

The paper is organised as follows. In Section \ref{sec:seaice}, we summarize relevant background on the mechanisms of sea ice melt pond evolution. In Section \ref{sec:TDA}, we introduce persistent homology and the TDA pipeline. In Section \ref{sec:methods}, we present and adapt the signed Euclidean distance transform \citet{RobinsSEDT}, and describe the dataset. In Section \ref{sec:results}, we present and discuss the results of our analysis and offer interpretations. We propose future directions for topological analysis of melt ponds in Section~\ref{sec:discussion}.

\section{\label{sec:seaice}Sea Ice Background}
The evolution of seasonal Arctic sea-ice melt ponds is described by \citet{SeaIceAlbedoModel}, who describe a 7-phase model of melt pond evolution throughout the year. Starting from late May, melt ponds begin appearing on the surface of the ice. Meltwater volume begins to increase until around early June, when water suddenly begins to drain to sea level. Mechanisms for this drainage are investigated in \citet{polashenskimechanisms}, which points to drainage through macroscopic ice flaws as a major factor. These can arise as breathing holes broken in the ice by seals, but \citet{polashenskimechanisms} conjectures that it is more likely that these are holes formed via the widening of pores in the ice as warmer water permeates through them.

After this rapid drainage, the water level within the ponds remains steady at roughly sea level, as the surrounding ice continues to melt. In the case of seasonal ice, eventually the ice melts completely by around August. Seawater begins to freeze again in mid-August, and new seasonal ice forms.

\section{\label{sec:TDA}Topological Data Analysis Background}

Homology is a classical topological invariant which allows us to quantify key properties of the shape of an object \cite{hatcher}. The \textit{0\textsuperscript{th} homology group} is a vector space whose dimension corresponds to the number of distinct connected components (``0-cycles") an object possesses. The \textit{1\textsuperscript{st} homology group} is a vector space whose dimension gives the number of topologically distinct loops (``1-cycles") in the object. Higher homology groups track voids and holes of larger dimension, though we do not compute them in this analysis. \textit{Persistent homology} tracks changes in homological features as a resolution parameter increases, which includes more and more of the underlying object. 

The key input for persistent homology is a \textit{filtration} on a simplicial complex. A simplicial complex is a set of vertices connected together by edges (1-simplices), triangles (2-simplices) and higher-dimensional simplices. A \textit{filtration} is a sequence of simplicial complexes such that each is contained in the next simplicial complex in the sequence. Intuitively, this corresponds to `building up' a simplicial complex by adding more simplices at each step. For each simplicial complex in the filtration, we can calculate the $n$\textsuperscript{th} homology, and by studying the maps induced on the homology groups by the inclusions between simplicial complexes, we can determine the filtration values at which a topological feature like a 0-cycle first appears (the `birth value') and disappears (the `death value') \cite{carlssonzomorodian}.

\begin{figure}[htp]
\subcapcentertrue
\centering
 \subfigure[2D Height Map]{\label{fig:sublevel-a}\includegraphics[width=0.4\columnwidth]{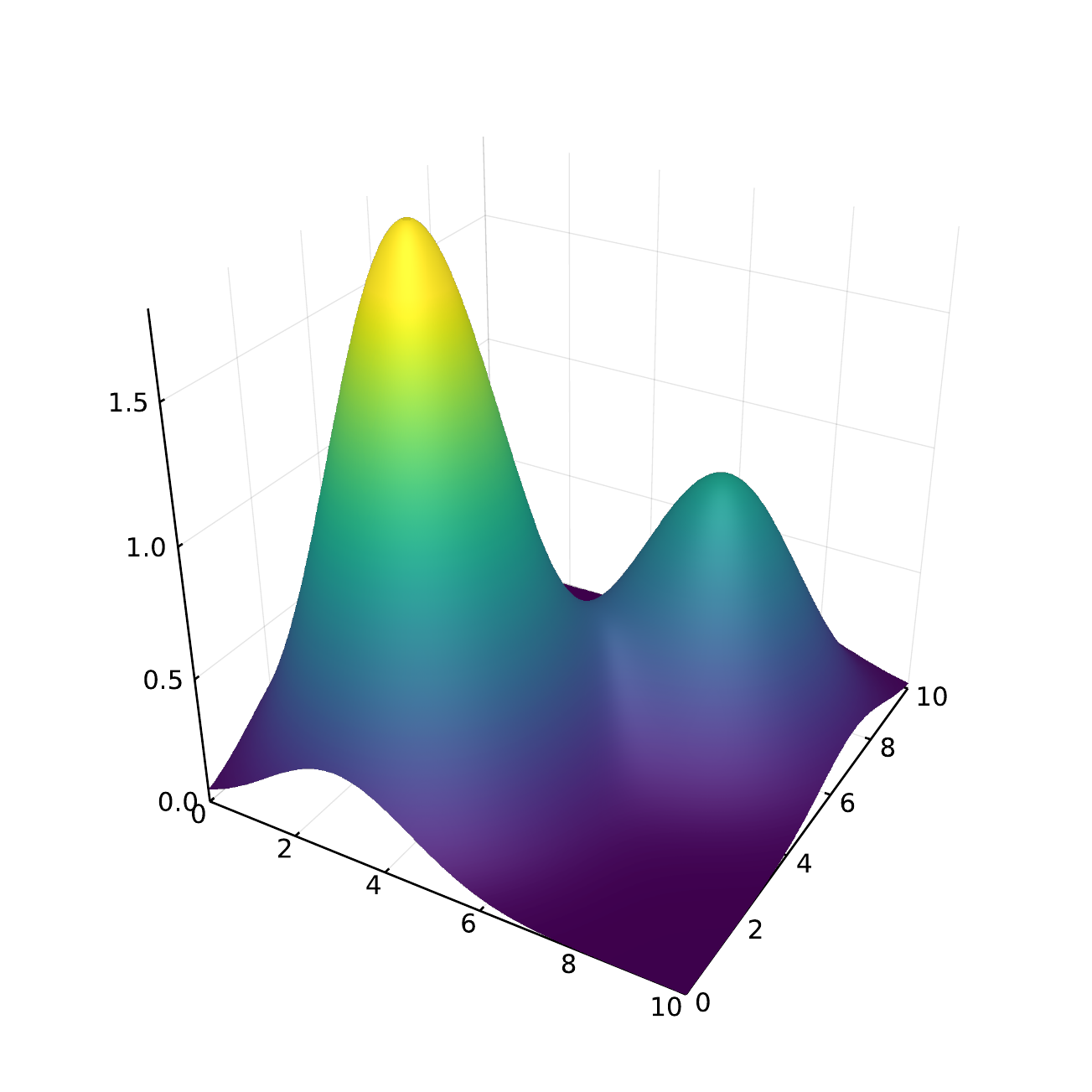}}%
%
   \subfigure[Sublevel Filtration]{\label{fig:sublevel-b}\vbox{\offinterlineskip\halign{#\hskip3pt&#\cr
  \includegraphics[width=0.2\columnwidth]{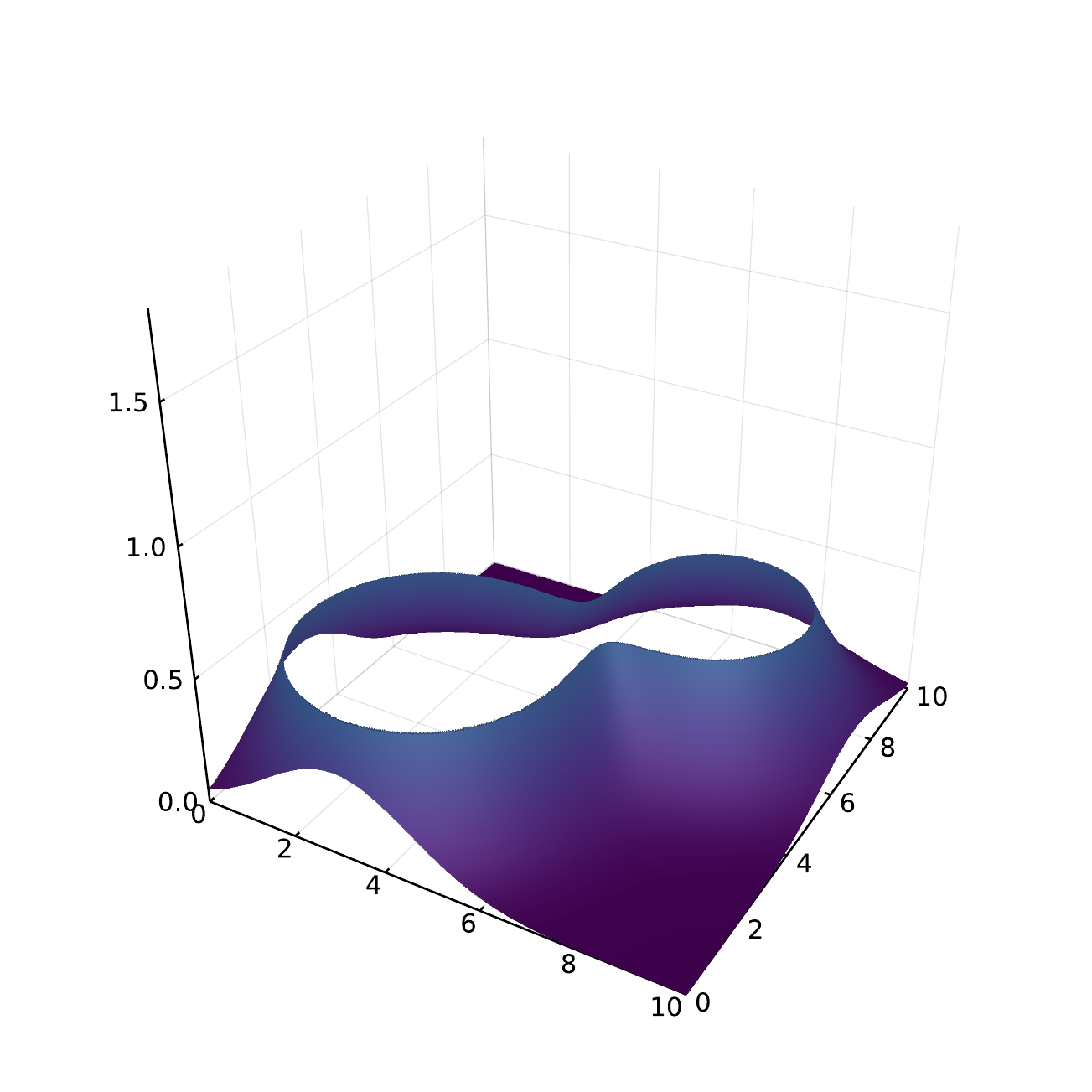}&
  \includegraphics[width=0.2\columnwidth]{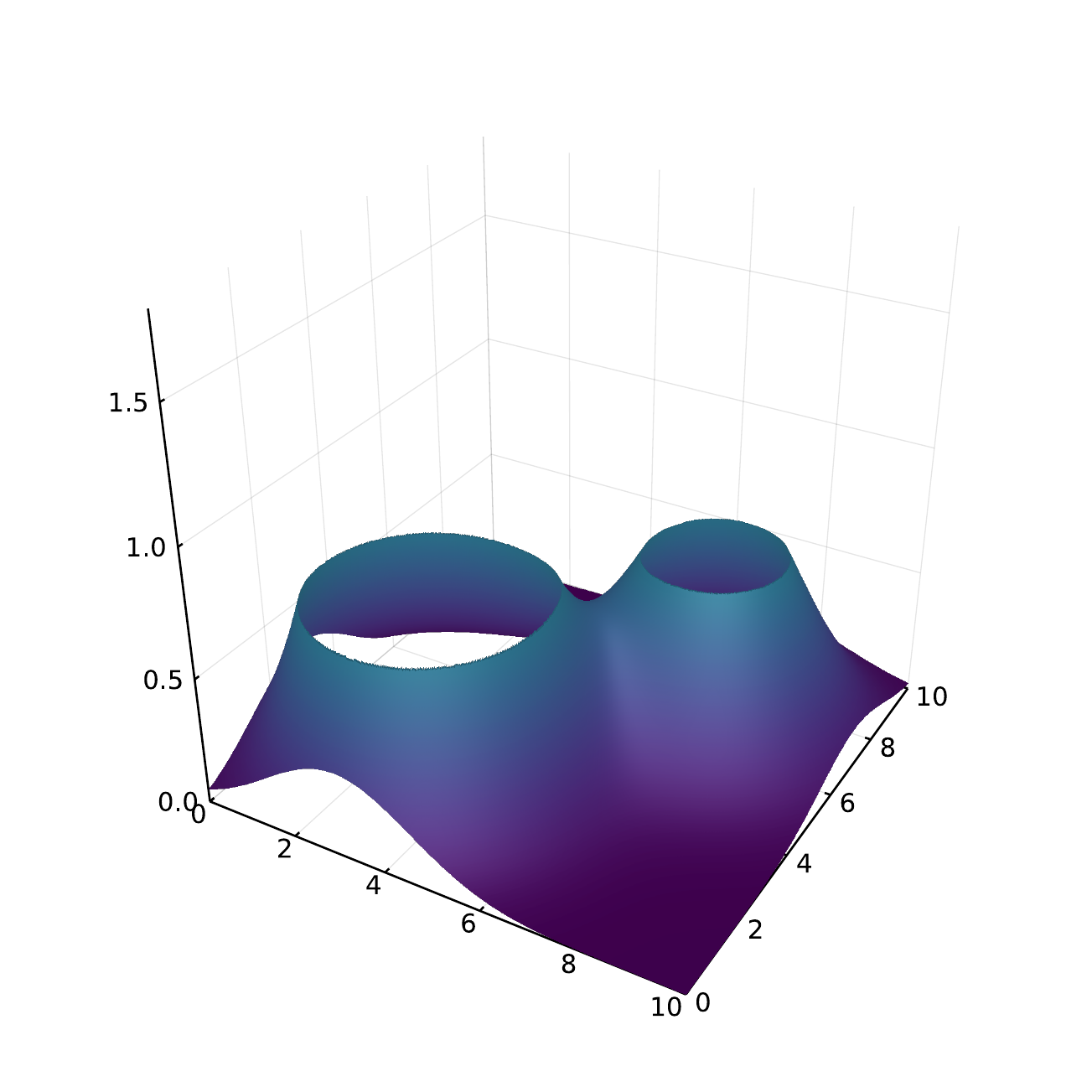}\cr
  \noalign{\vskip3pt}
  \includegraphics[width=0.2\columnwidth]{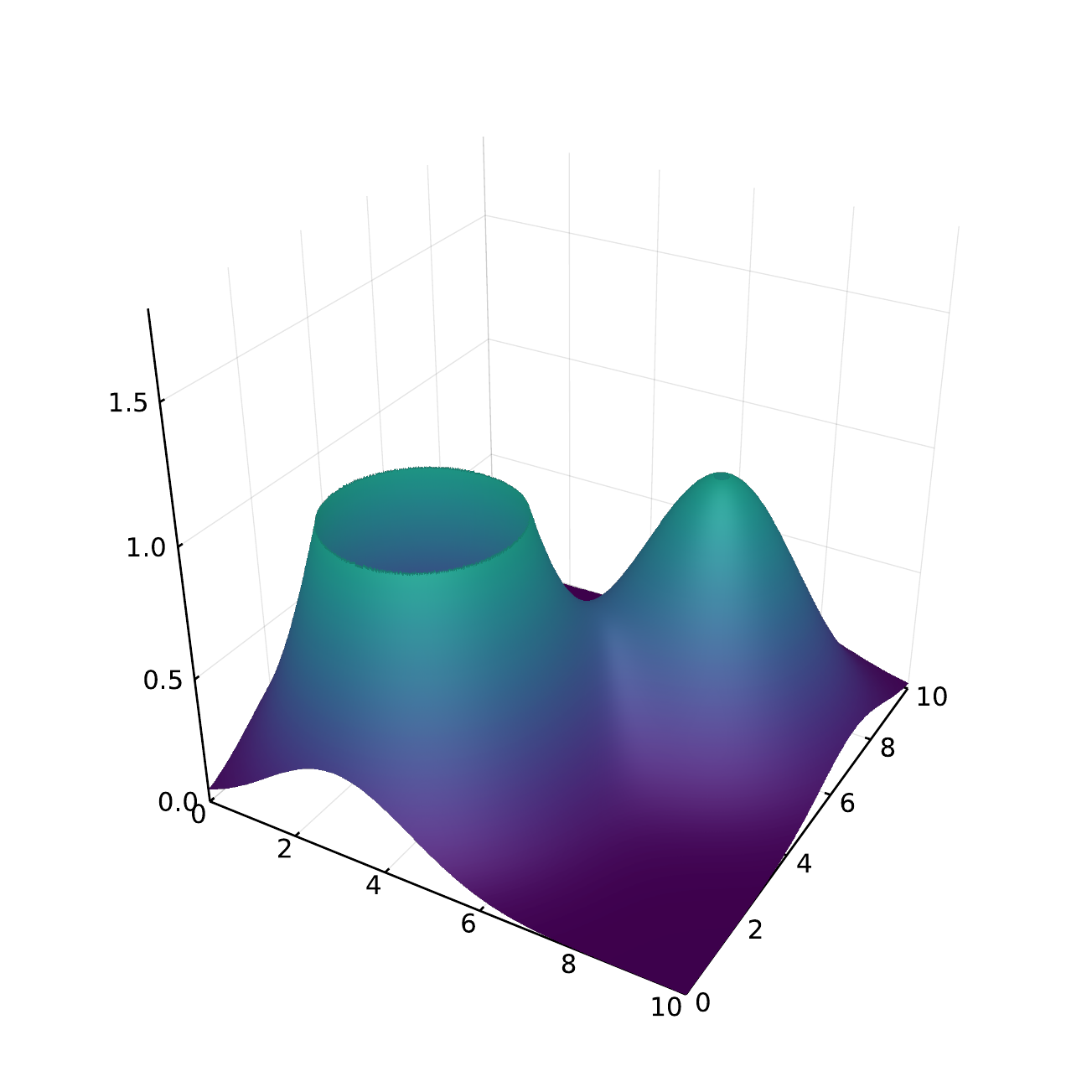}&
  \includegraphics[width=0.2\columnwidth]{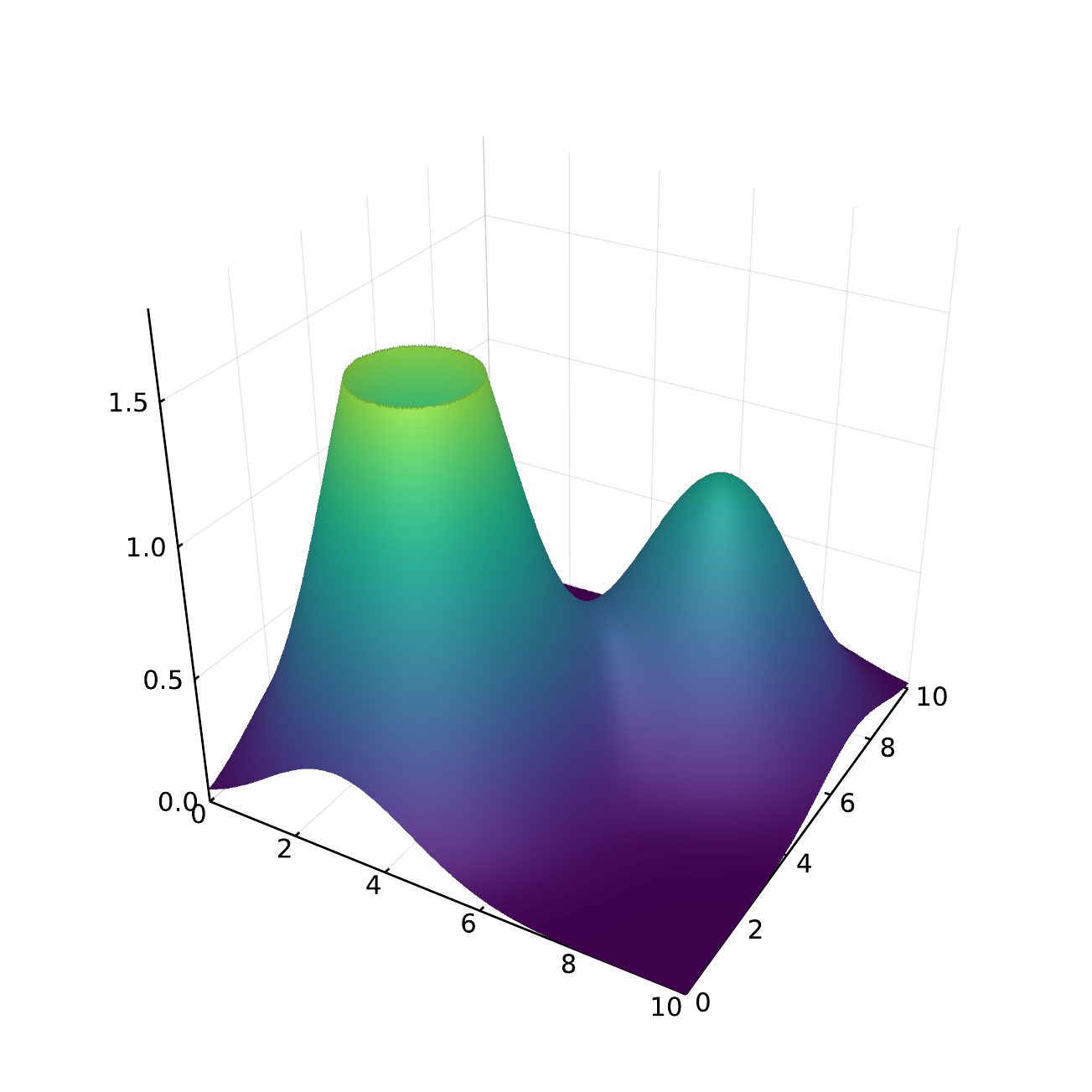}\cr
  }}}

 \subfigure[Persistence Diagrams]
 {\label{fig:sublevel-c}\includegraphics[width=0.4\columnwidth]{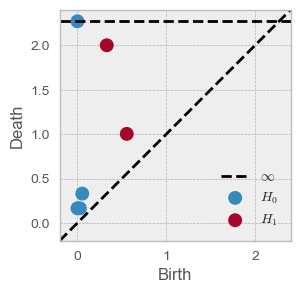}}

\label{fig:sublevel}
\caption{\centering The sublevel filtration and persistence diagrams for a function on $\mathbb{R}^{2}$. The red points represent the two persistent 1-cycles (`holes'), their birth at saddle points and critical values of the boundary, and their death values at local maxima.}
\end{figure}

Concretely, suppose we have the graph of a function represented as a 2D height map lying above the plane (Fig.~\ref{fig:sublevel-a}). For each value of height $h$, we can form a {\em sublevel set} simplicial complex including only those pixels which lie below height $h$, along with their immediate adjacencies. Taking the range of possible $h$ values produces a filtration (Fig.~\ref{fig:sublevel-b}). By computing homology for each step along the filtration, we can track the heights at which holes and connected components first appear (``birth" value) and when they merge into surrounding features (``death"). The persistent features contain the critical values of $h$: a local maximum at height $h_a$ corresponds to a hole which closes at step $h_a$, so there is a generator of $H_1$ with death value $h_a$, and similarly a local minimum at height $h_b$ will produce a new connected component in $H_0$ born at height $h_b$. Plotting the (birth height, death height) pairs, we obtain a \textit{persistence diagram} (Fig.~\ref{fig:sublevel-c}). 
We call the (death-birth) lifespan of a feature its {\em persistence}; points far from the diagonal represent the most persistent features. Highly persistent points correspond to local maxima or minima that stand out in high contrast to their surrounding area.

\section{\label{sec:methods}Methods}
\begin{figure}[htp]
  \begin{center}
    \includegraphics[width=0.4\columnwidth]{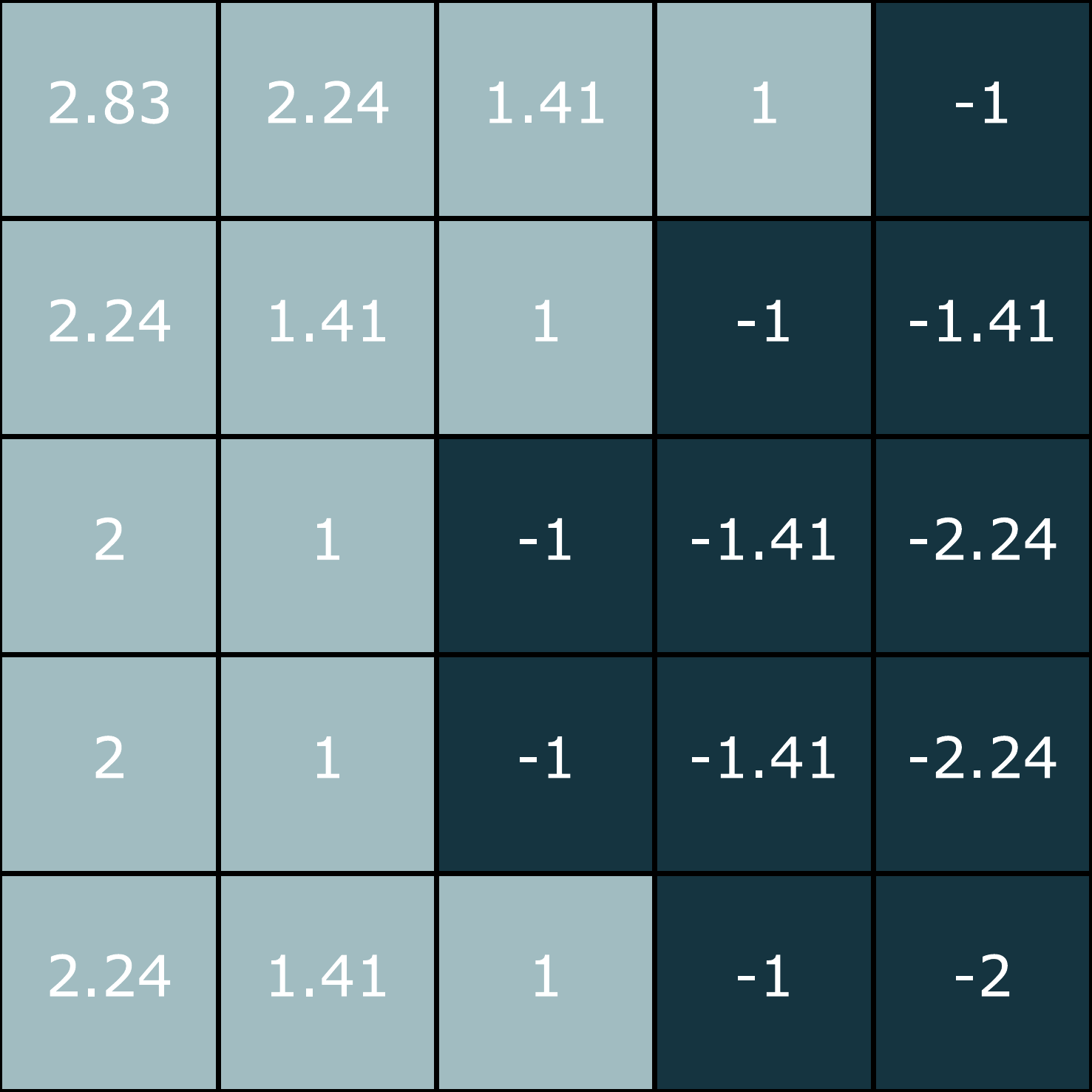}
    
  \end{center}
  \caption{Signed Euclidean Distance Transform of a Binary Image}
  \label{fig:SEDTexample}
\end{figure}

We start with a topological method proposed by \citet{RobinsSEDT} to analyse binary images. This approach constructs a filtration that encodes  complementary geometry and connectivity of the two phases of the binary partition, and then applies persistent homology. For each pixel of the image, we calculate the distance to the nearest boundary between phases (more precisely, the distance to the nearest pixel of the opposite phase). We then assign this distance value to the pixel, along with a sign encoding in which phase of the material the pixel lies. For instance, a water pixel which is $3$ units away from the nearest ice will be assigned a value of $-3$, while an ice pixel which is 3 units away from the nearest water pixel will be assigned a value of $+3$. This function is called the signed Euclidean distance transform, or SEDT (Fig.~ \ref{fig:SEDTexample}).

From the SEDT values, we construct a sublevel set filtration of the underlying lattice. Applying persistent homology to the resulting sublevel filtration, we obtain persistence diagrams for the 0\textsuperscript{th} and 1\textsuperscript{st} homology. The key observation is that birth and death times in each quadrant of the resulting persistence diagrams have different interpretations that are relevant for the connectivity and percolation of the binary image. For instance, negative birth values and positive death values correspond to features that can be seen in the original image. The other quadrants encode information about concavity of the phase regions. The interpretations are listed in Table \ref{tab:SEDTInterpretation}. For a similar table interpreting birth/death values for 3D binary structure in porous materials, see \citet{statisticalinference}.

\begin{table*}[htp]

\caption{\label{tab:SEDTInterpretation}Interpretations of persistence diagram for sublevel SEDT filtration}
\centering
\includegraphics[width=\textwidth]{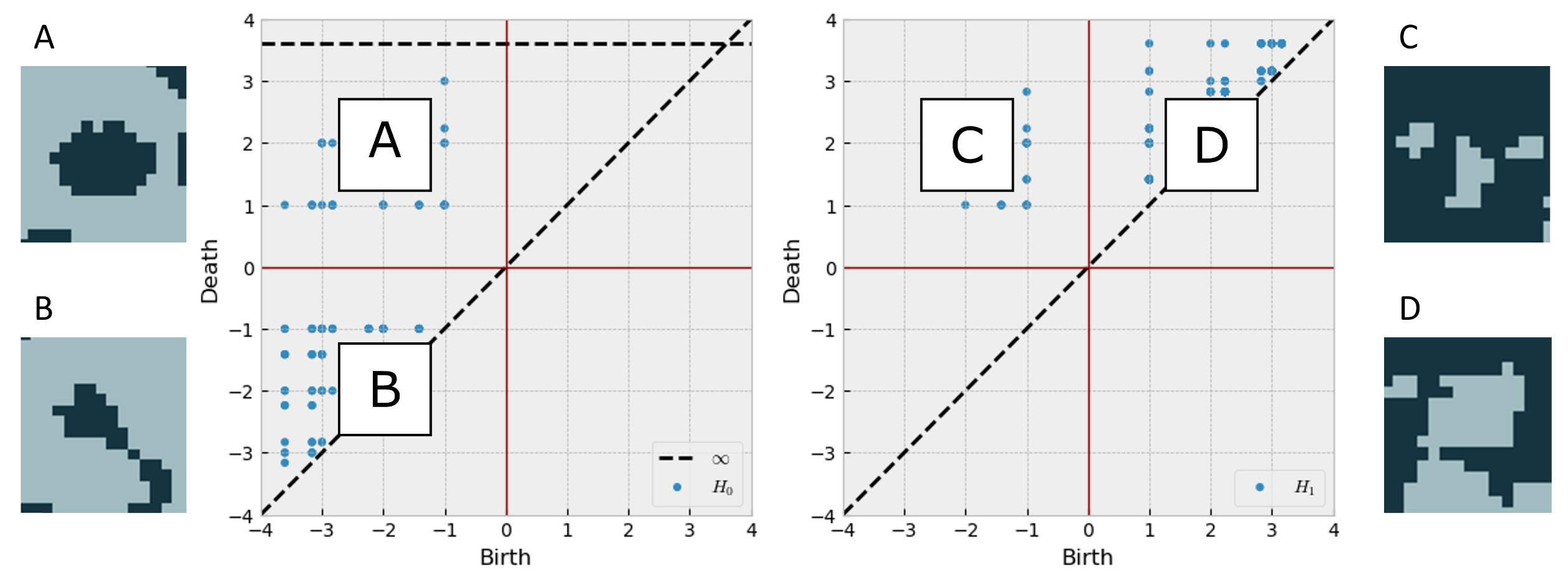}
\renewcommand{\arraystretch}{2}

\begin{ruledtabular}
\begin{tabular}{cp{0.2\textwidth}p{0.2\textwidth}p{0.2\textwidth}}
Region & Interpretation & Birth Values & Death Values\\ \hline
A & Pond Connected \newline Components & Pond Radius & Separation from \newline other Ponds\\
B & Node-Like Subponds & Nodal Pond Radius & Connecting Channel\newline Throat Radius\\
C & Pond Loops & Pond Loop \newline Throat Radius & Radius of Ice \newline Enclosed in Loop\\
D & Node-Like Subregions of \newline Concave Islands of Ice & Radius of Throat of \newline Concave Ice Island & Concave Ice Island \newline Node Radius
\end{tabular}
\end{ruledtabular}

\end{table*}

\begin{figure}[h]
  \begin{center}
    \subfigure[Time step 10]{\label{fig:model-a}\includegraphics[width=0.3\columnwidth]{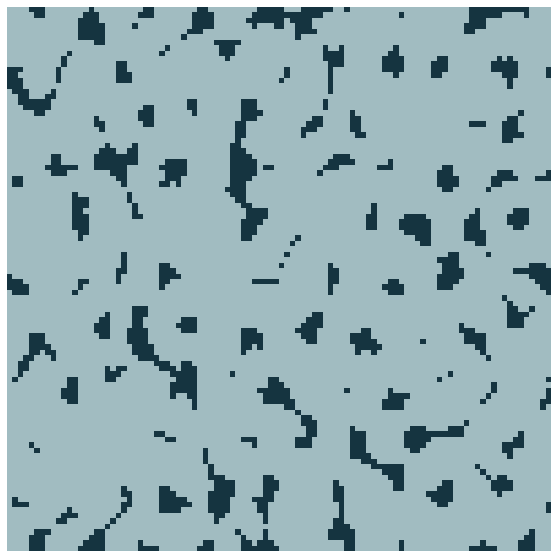}}
    \subfigure[Time step 100]{\label{fig:model-b}\includegraphics[width=0.3\columnwidth]{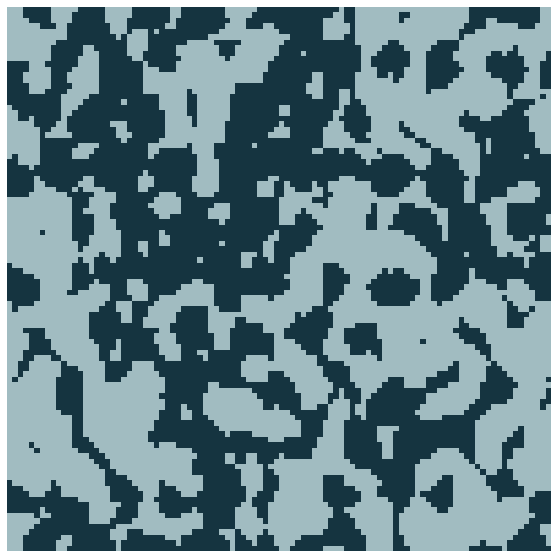}}
    \subfigure[Time step 200]{\label{fig:model-c}\includegraphics[width=0.3\columnwidth]{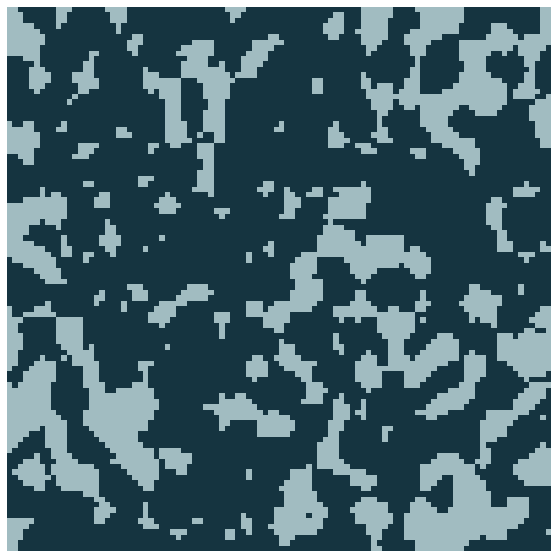}}
  \end{center}
  \caption{Output from a single run of the first dynamical network model.}
  \label{fig:model}
\end{figure}

 We apply the above method to a synthetic set of time-series of binary images coming from a dynamical network model for melt pond evolution. Full details of the model are available in the DPhil thesis 
 of \citet{coughlan2021dynamical}. 
 We have three variations of the model: 
 \begin{itemize}
     \item {\em Model 1} uses a constant melt rate
     \item {\em Model 2} has melt rate varying depending on albedo
     \item {\em Model 3} introduces pond drainage via randomly occurring drainage events
 \end{itemize} For each variation, we generate 10 runs of the model, and produce 201 time steps of binary images depicting pond evolution (Fig.~ \ref{fig:model}). We then calculate the SEDT of each binary image, assigning pond pixels negative values and ice pixels positive values. We compute the persistence diagrams for the sublevel filtration on the SEDT using the software package Ripser, and then extract a list of $(\text{birth}, \text{death})$ pairs in each quadrant of the persistence diagrams for 0\textsuperscript{th} and 1\textsuperscript{st} homology. This allows us to extract the statistics listed in Table~\ref{tab:SEDTInterpretation}, such as number of disconnected ponds, and average pond radius.
 
\begin{figure}[h]
  \begin{center}
    \subfigure[]{
     \includegraphics[height=0.25\columnwidth]{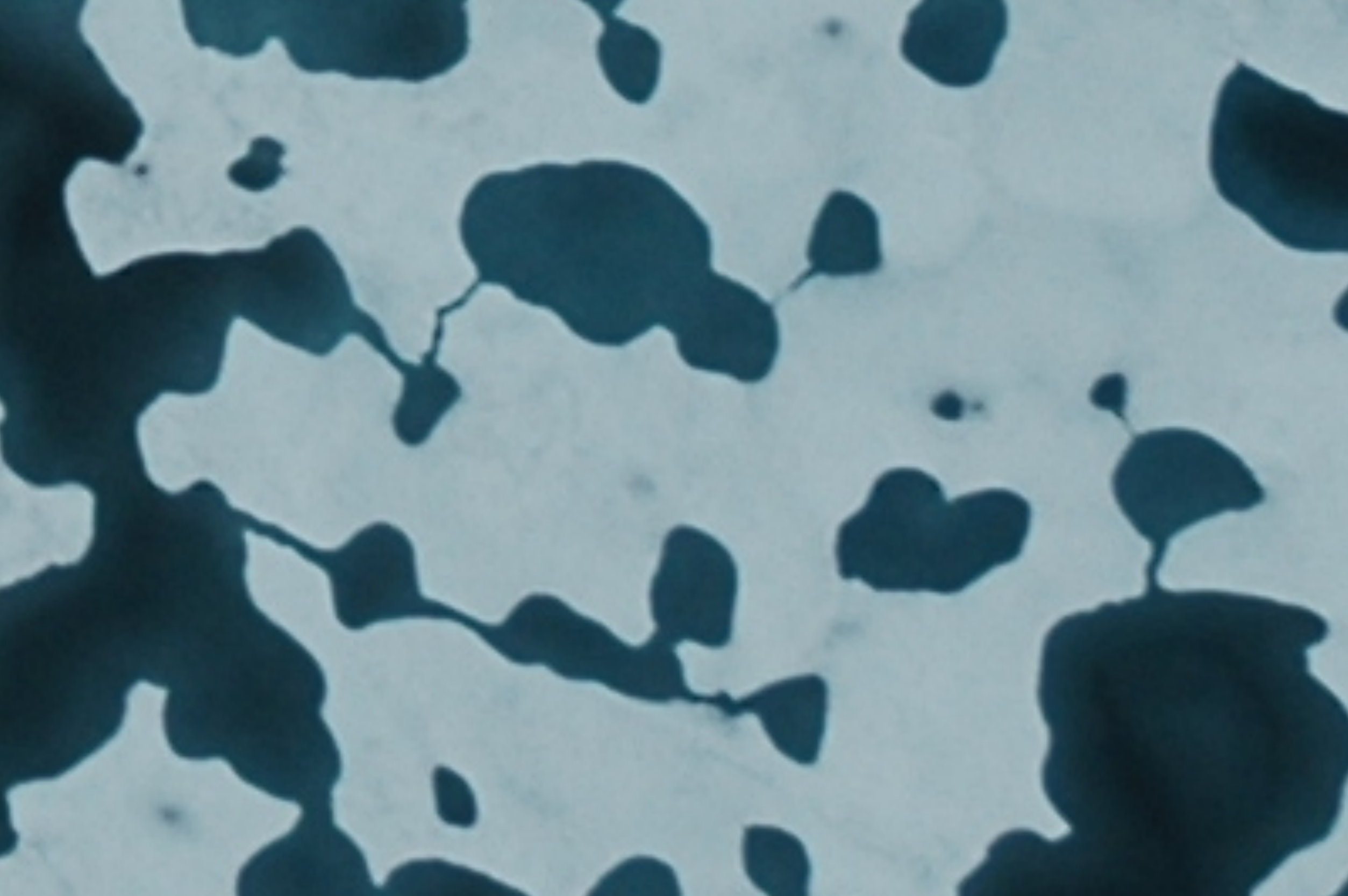}}
  \subfigure[]
         {\includegraphics[width=0.25\columnwidth]{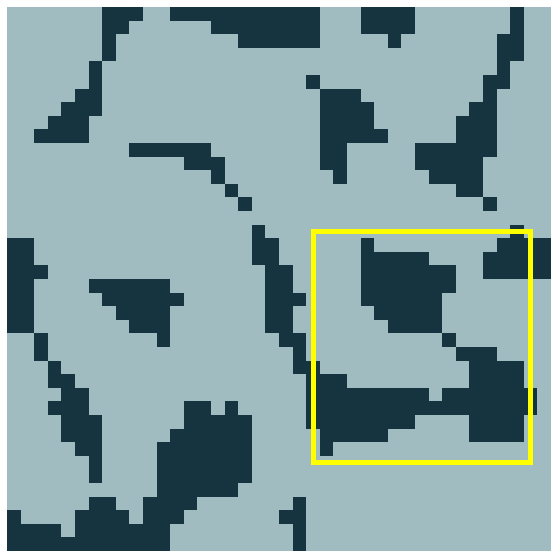}}
    \includegraphics[width=0.25\columnwidth]{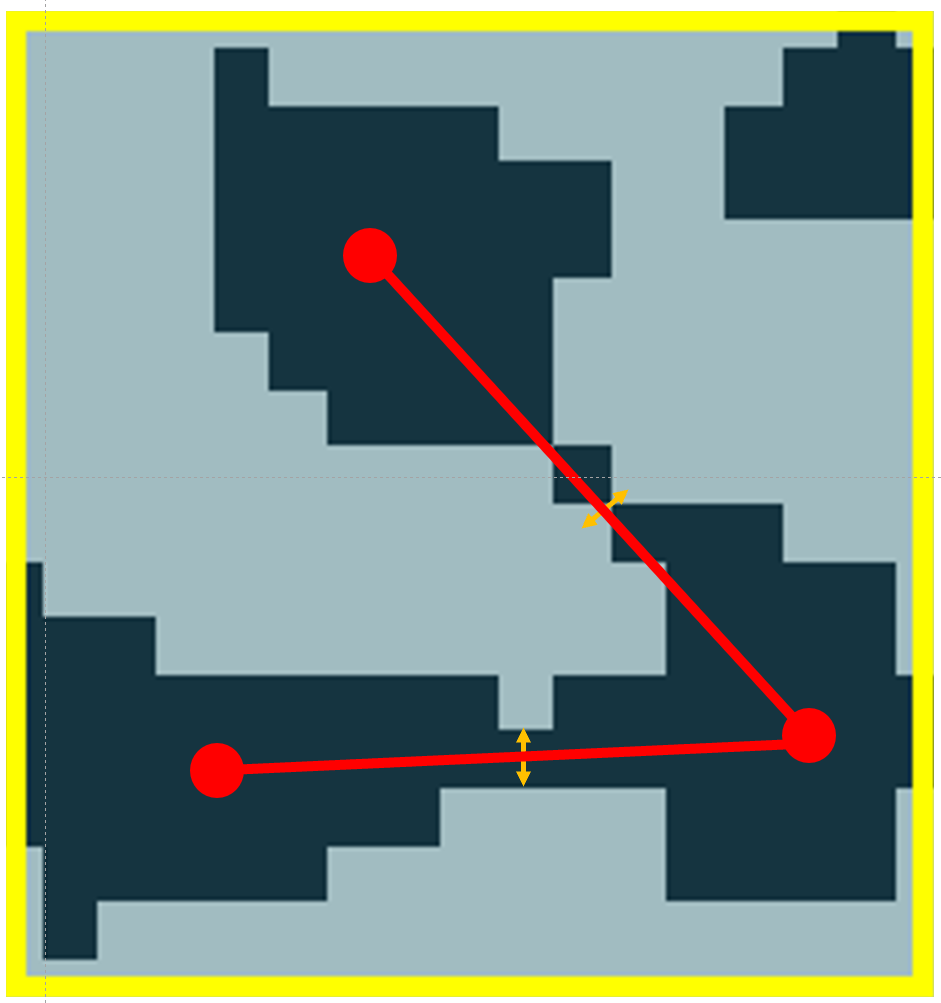}

  \end{center}
  \caption{\centering Regions of pond showing network-like structure in both (a) real melt ponds, and (b) simulated melt ponds. Circular `nodal' subponds are connected by edge-like chanels with narrow throats, shown enlarged at right. Credit: Perovich for photograph}
  \label{fig:graph structure}
\end{figure}

 In addition to those listed in Table \ref{tab:SEDTInterpretation}, some other topological statistics are of interest. Pond evolution often leads to `node/edge' structures arising when two or more more-or-less circular ponds connect to one another by a channel between them (Fig.~\ref{fig:graph structure}). Birth-death pairs $(b,d)$ in quadrant B (i.e. $b,d<0$) correspond to these node/edge structures, where $-b$ is the radius of the smaller pond and $-d$ is the throat radius of the connecting channel. The persistence $d-b$ of such a pair then tells us the `aspect-ratio' of such a structure, giving a measure of its concavity.
 
We also investigate percolation thresholds for melt pond evolution. Critical area thresholds for percolation have been studied for melt ponds using fractal dimension in \citet{FractalDimensionPonds}, \citet{Ising}, \citet{simplerules}. We investigate another way of determining percolation. Following \citet{RobinsSEDT}, we define the percolation threshold $r_{perc}$ to be the radius of the smallest sphere that can freely pass through the pond phase from one side of the image to another. We investigate this for melt ponds in two complementary ways.

Firstly, following \citet{flowestimation}, we can consider 0-holes in the ice phase, i.e. pond channels which divide the ice into two or more connected components. Note that a connected component of ice can arise in one of two ways. First, a channel spanning the whole image will divide the ice into two components. Second, an internal `loop' in a pond will section off an island of ice within it. As in \citet{flowestimation}, we aim to subtract ice components arising in the second way from the set of components arising in the first way.

 To compute these components, we can take a \textit{superlevel} filtration of the signed Euclidean distance function, where we only include pixels \textit{above} a given height. Region A of the 0\textsuperscript{th}-homology persistence diagram then tells us information about connected components of the ice, rather than the ponds. In particular, given a birth-death pair $(b,d)$, the death value $d$ tells us the thinnest throat radius of the pond separating an ice component from its nearest neighbour. If the ice component has arisen as an island encircled by a pond loop, the death value $-d$ will also arise as the \textit{birth} value of a birth-death pair in region C of the 1\textsuperscript{st} homology of the sublevel filtration. The remaining technical details are addressed in Section B of the Supplementary Material. We can then filter out all values that also occur in this region, and are left with a list containing only the throat radii of pond channels which span the entire image.
 
\begin{figure}[h]
  \begin{center}
     \includegraphics[width=0.25\columnwidth]{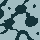}
  \end{center}
  \caption{\centering Channels which are short but span from edge to edge of the image: this arises in the case of channels which cross the diagonal of a corner (top-left) and those which begin and end on the same side of the image boundary (bottom).}
  \label{fig:small spanning}
\end{figure}

 The above method is useful when there are relatively few `entrance' and `exit' points of the water phase, as in the fracture models considered in \citet{flowestimation}. In some cases, however, we have many short channels which nevertheless span from boundary to boundary of an image. This happens when a channel begins and ends on the same edge of the image, or where a channel spans across the diagonal of a corner of the image (Fig.~\ref{fig:small spanning}). In scenarios where these short spanning channels are common, we may wish instead to detect pond channels which connect opposite edges of an image.

 With a slight variation on this technique, we can compute the value $r_{perc}$, the minimum throat radius of channels which span from one side of the image to the opposite side. If this value is negative, it means opposite sides are not connected, and instead $r_{perc}$ measures the largest radius of ice which must be crossed to traverse the image from one side to the opposite. Details of this computation are included in the supplementary material. 
 This method has the benefit that we only detect channels which fully connect opposite sides of the image, which is what we want to work out when percolation begins. 

The code for the project is available at \href{https://github.com/wilfofford/TDA-for-Sea-Ice-Percolation}{https://github.com/wilfofford/TDA-for-Sea-Ice-Percolation}.

\section{\label{sec:results}Results and Interpretation}

\begin{figure*}[htp]
\subcapcentertrue
  \begin{center}
    \subfigure[Average Pond Radius]{\label{fig:SEDToutput-a}\includegraphics[width=0.3\textwidth]{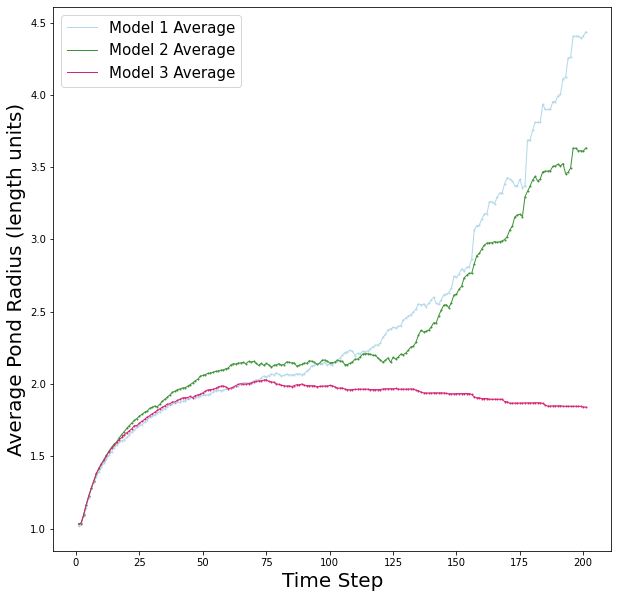}}
    \subfigure[Number of Ponds]{\label{fig:SEDToutput-b}\includegraphics[width=0.3\textwidth]{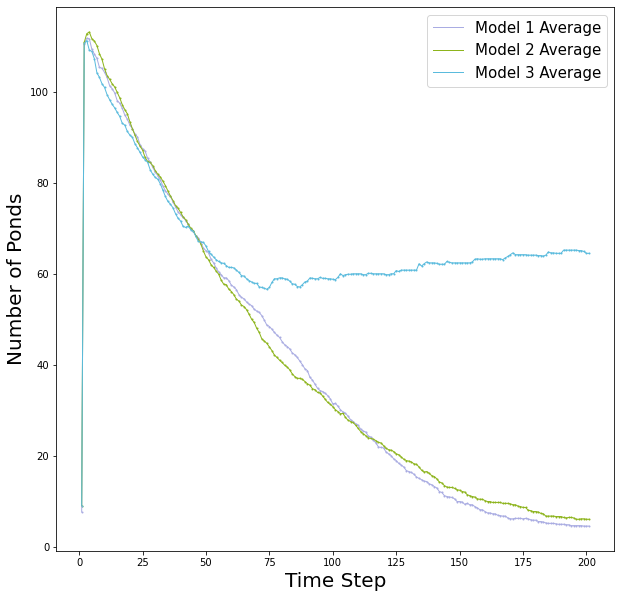}}
    \subfigure[Concave Pond Radius against Pond Coverage Fraction]{\label{fig:SEDToutput-c}\includegraphics[width=0.3\textwidth]{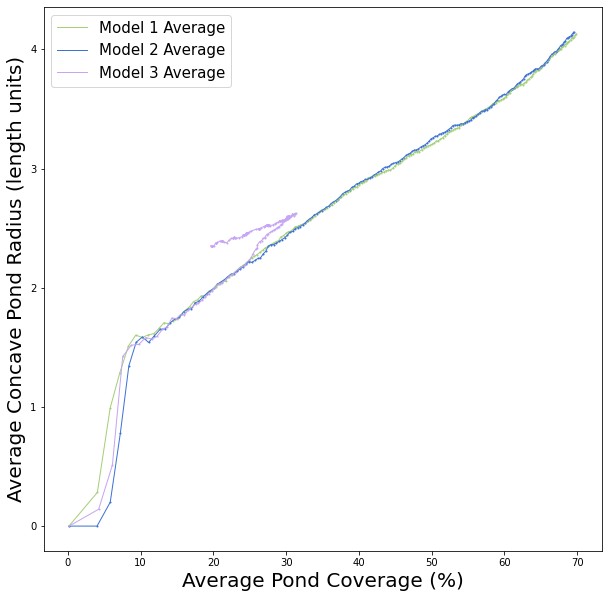}}\\
    \subfigure[Average Pond Aspect-Ratio against Time]{\label{fig:SEDToutput-d}\includegraphics[width=0.45\textwidth]{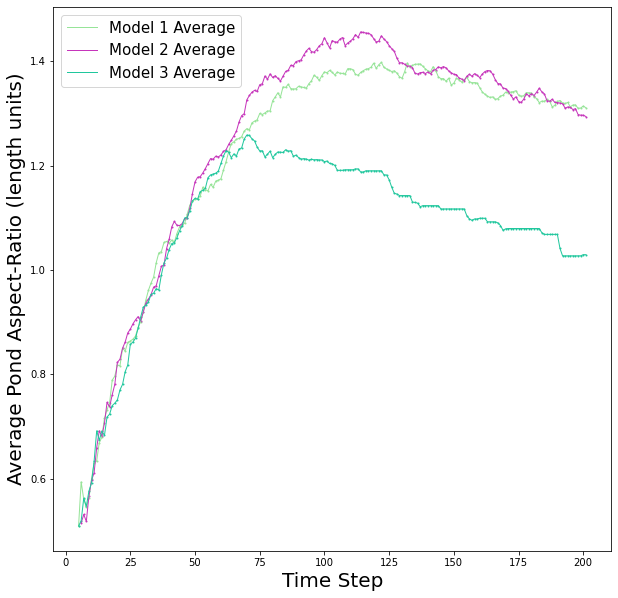}}
    \subfigure[Average Pond Aspect-Ratio against Pond Coverage Fraction]{\label{fig:SEDToutput-e}\includegraphics[width=0.45\textwidth]{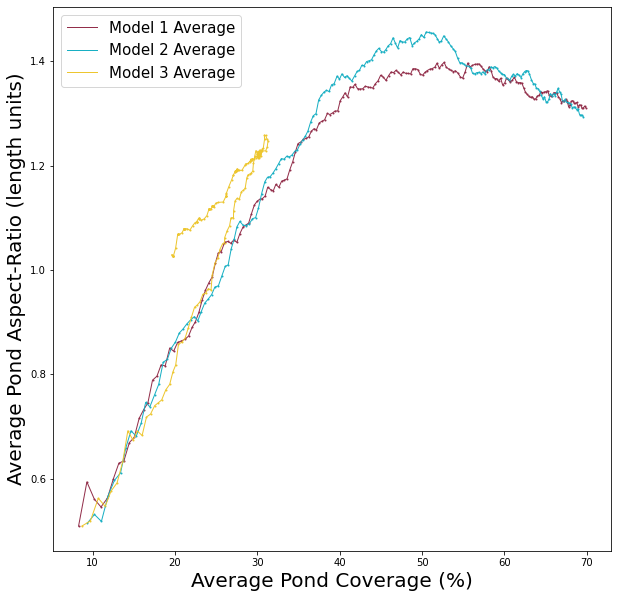}}
  \end{center}
  \caption{Persistence statistics of Signed Euclidean Distance Filtration for model data}
  \label{fig:SEDToutput}
\end{figure*}

Topological and geometric statistics arising from the SEDT are shown in Fig.~ \ref{fig:SEDToutput}. Plots for average pond radius (birth values in region A), number of ponds (number of birth-death pairs in region A) and average pond aspect-ratio (persistence of birth-death pairs in region B) are plotted against time, and we also plot the average aspect ratio and pond radius of concave ponds (birth values in region B) against the pond coverage fraction.

\begin{figure}[htp]
\subcapcentertrue
  \begin{center}
    \subfigure[Model 1, Run 7]{\label{fig:concaveponds-a}\includegraphics[width=0.3\columnwidth]{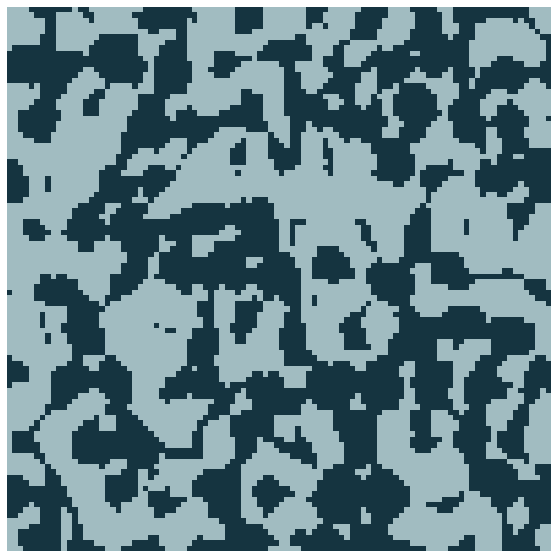}}
    \subfigure[Model 2, Run 3]{\label{fig:concaveponds-b}\includegraphics[width=0.3\columnwidth]{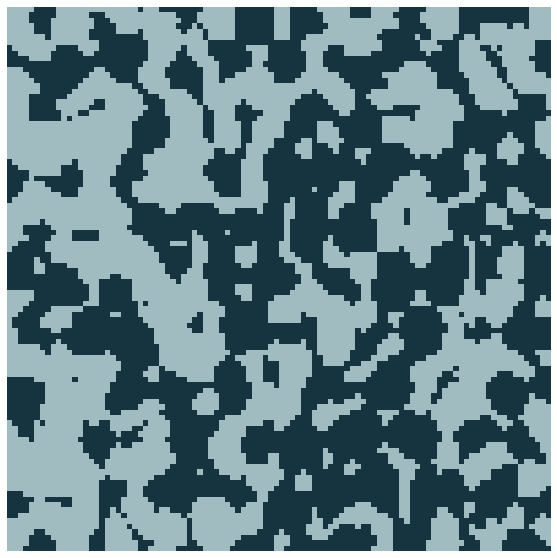}}
    \subfigure[Model 3, Run 3]{\label{fig:concaveponds-c}\includegraphics[width=0.3\columnwidth]{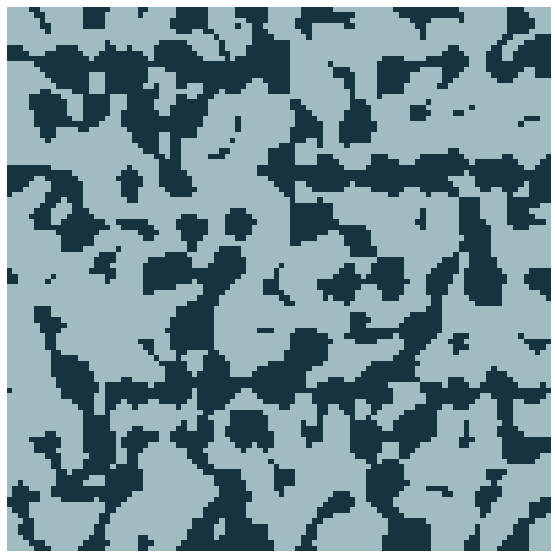}}
  \end{center}
  \caption{\centering Model runs with highest average aspect-ratio for each of the three models. The most concave images occur at run 7 timestep 102 of model 1, run 3 timestep 115 of the second model, and run 3 timestep 73 of model 3. These images display pronounced `network-like' structure, with circular nodal ponds linked by thin channels.}
  \label{fig:concaveponds}
\end{figure}

From the average pond aspect-ratio in Fig.~\ref{fig:SEDToutput-d}, we see that all 3 models follow a similar trend: aspect-ratio rapidly increases up to a maximum, and then gradually decreases again. Since pond evolution often forms patterns of rounder `nodal' ponds connected via thinner channels, we can interpret the aspect-ratio as a measure of the relative sizes of the nodal ponds and channels between them. Aspect-ratio should increase as the network-like structure becomes more pronounced and the ponds more concave, explaining the initial increase. After a certain point, however, in Models 1 and 2, subsequent ice melt results in the linking channel radius increasing faster than the nodal pond radius, and hence aspect-ratio decreases. Model 2 achieves a noticeably higher aspect-ratio, indicating that this model results in more highly-pronounced network-like pond structure. The most concave pond images for each model are shown in Fig. \ref{fig:concaveponds}.
In Model 3, drainage halts the increase of meltwater contained in the ponds, preventing new connections from forming and reducing the average aspect-ratio of the network.


The pond radius and pond number, Figs. \ref{fig:SEDToutput-a} and \ref{fig:SEDToutput-b}  show that while Models 1 and 2 produce a similar quantity of ponds through time, Model 1 produces on average wider ponds in later stages of melt.
 For Model 3, which adds pond drainage to the variable melt rate, Figs.~\ref{fig:SEDToutput-c}-\ref{fig:SEDToutput-e}, average concave pond radius and average aspect-ratio, show a bifurcation point when drainage begins. Therefore, in Model 3, there is a difference in these topological statistics for ponds before and after drainage, even though they may have the same pond area. This suggests the possibility of classifying melt ponds by melt stage just from image data topology, although more work needs to be done to produce a robust classification method that works for real pond images.

\begin{figure*}[htp]
\subcapcentertrue
  \begin{center}
    \subfigure[Number of spanning channels against time]{\label{fig:spanning-a}\includegraphics[width=0.3\textwidth]{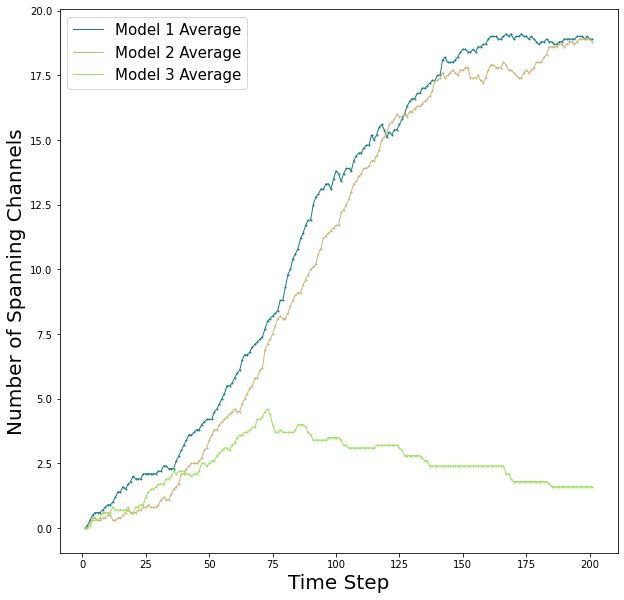}}
    \subfigure[Average spanning channel throat radius against time]{\label{fig:spanning-b}\includegraphics[width=0.3\textwidth]{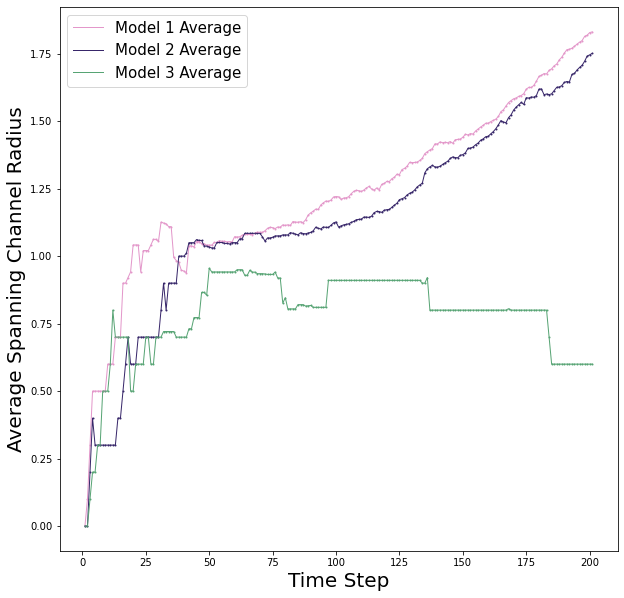}}
    \subfigure[Number of spanning channels against average pond radius]{\label{fig:spanning-c}\includegraphics[width=0.3\textwidth]{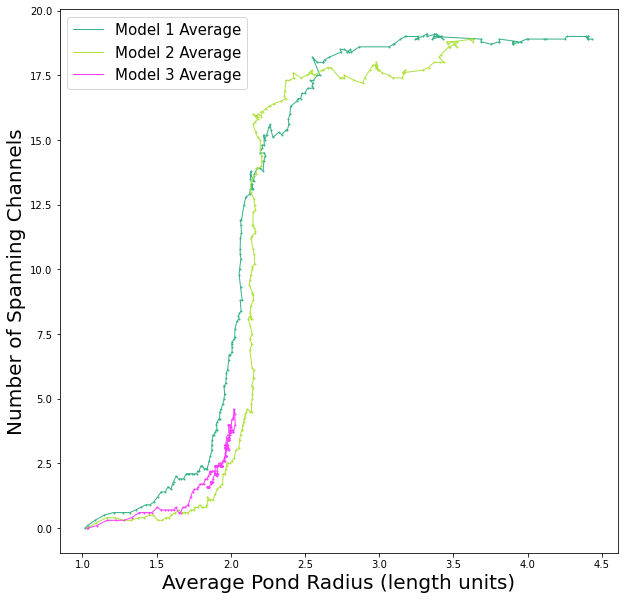}}
  \end{center}
  \caption{\centering Results for computation of spanning channels via superlevel set persistence for model data.}
  \label{fig:spanning}
\end{figure*}

We now move on to results relating to the percolation behaviour of the model. In Fig.~\ref{fig:spanning}, spanning channels are computed with a superlevel set filtration. In Fig.~\ref{fig:spanning-a}, we see that the number of channels which run from edge to edge follows a roughly sigmoidal shape for Models 1 and 2, while the drainage model has far fewer such channels. Most strikingly, in Fig.~\ref{fig:spanning-c}, we see a sharp transition in the number of spanning channels as the average pond radius increases, indicating the existence of a critical length scale of around 2.1 units of length, above which average pond radius percolation begins. From this graph also we can measure how close Model 3 is to attaining this critical length threshold; we can see that the pink line is just starting to creep up the sharp cliff-edge traced by the other two models. 

\begin{figure*}[htp]
\subcapcentertrue
  \begin{center}
    \subfigure[$r_{perc}$ against time for model 1]{\label{fig:percolation-a}\includegraphics[width=0.25\textwidth]{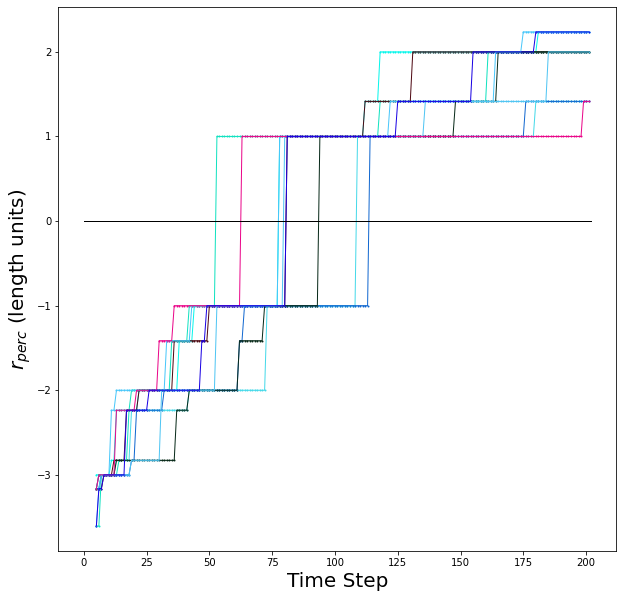}}
    \subfigure[$r_{perc}$ against time for model 2]{\label{fig:percolation-b}\includegraphics[width=0.25\textwidth]{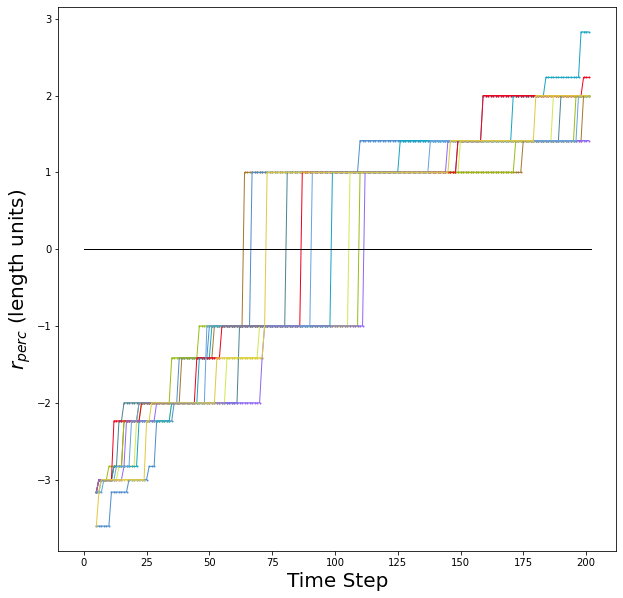}}
    \subfigure[$r_{perc}$ against time for model 3]{\label{fig:percolation-c}\includegraphics[width=0.25\textwidth]{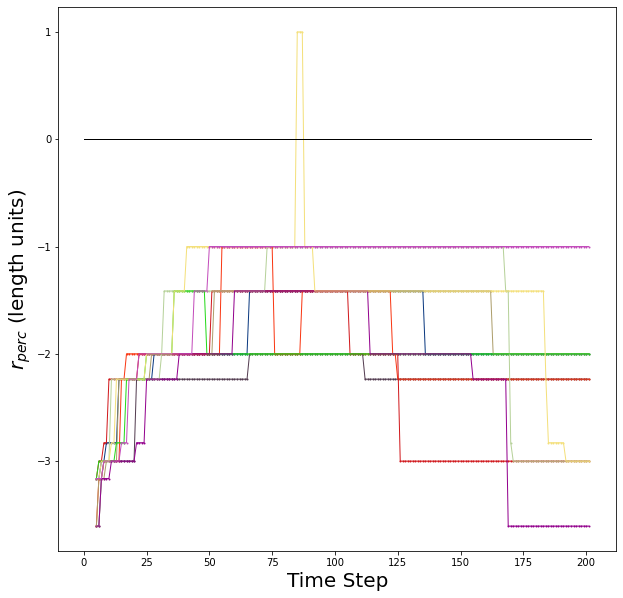}}
    \subfigure[Average $r_{perc}$ against time for each model]{\label{fig:percolation-d}\includegraphics[width=0.25\textwidth]{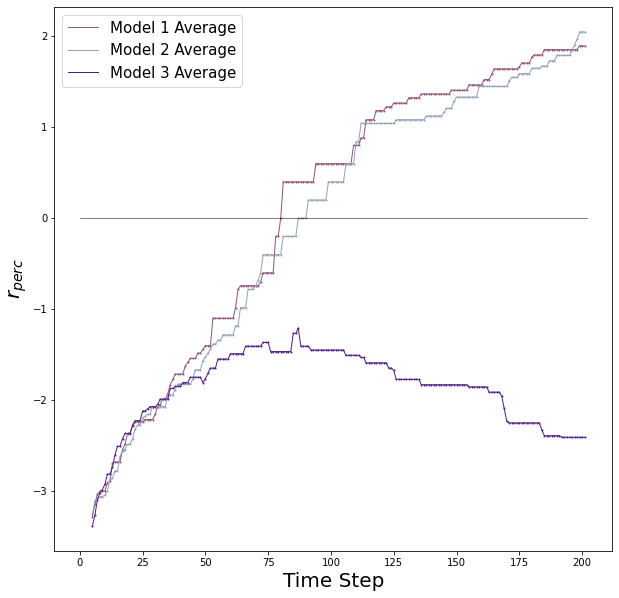}}
    \subfigure[Average $r_{perc}$ against pond coverage fraction for each model]{\label{fig:percolation-e}\includegraphics[width=0.25\textwidth]{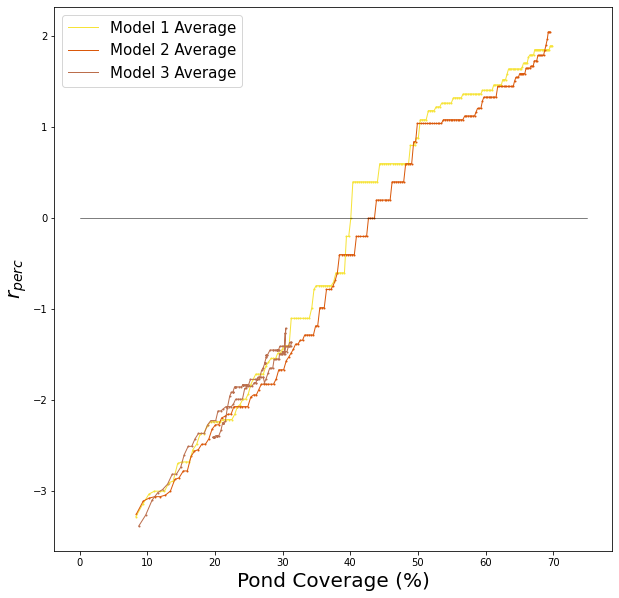}}
    \subfigure[Average $r_{perc}$ against average pond radius for each model]{\label{fig:percolation-f}\includegraphics[width=0.25\textwidth]{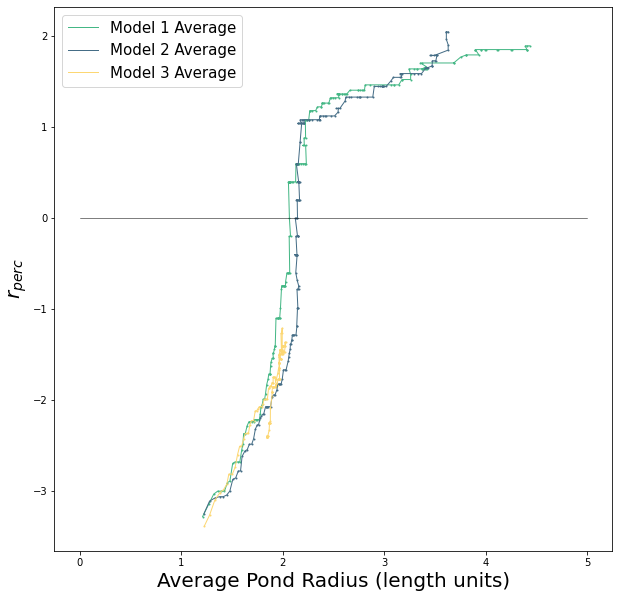}}
    \subfigure[Box plots for time at which opposite edges are connected]{\label{fig:percolation-g}\includegraphics[width=0.25\textwidth]{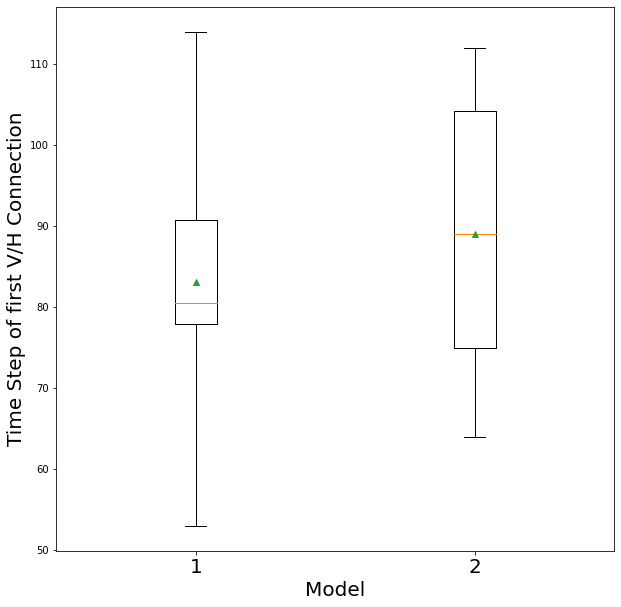}}
    \subfigure[Box plots for pond coverage fraction when opposite edges first connect]{\label{fig:percolation-h}\includegraphics[width=0.25\textwidth]{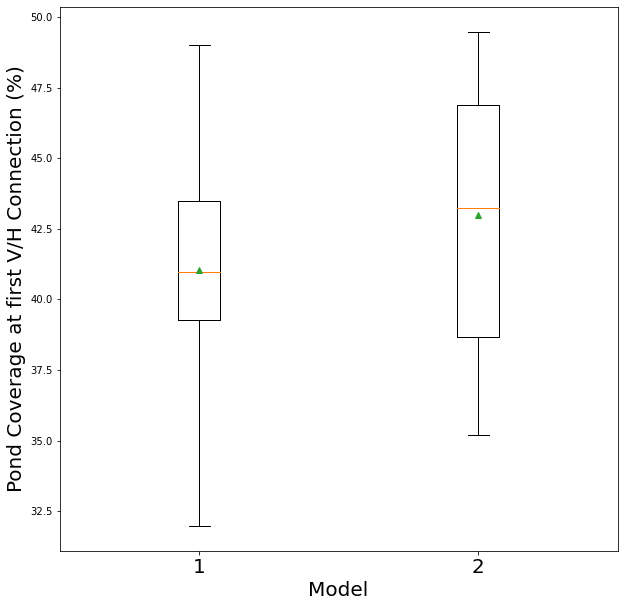}}
    \subfigure[Box plots for average pond radius at which opposite edges first connect]{\label{fig:percolation-i}\includegraphics[width=0.25\textwidth]{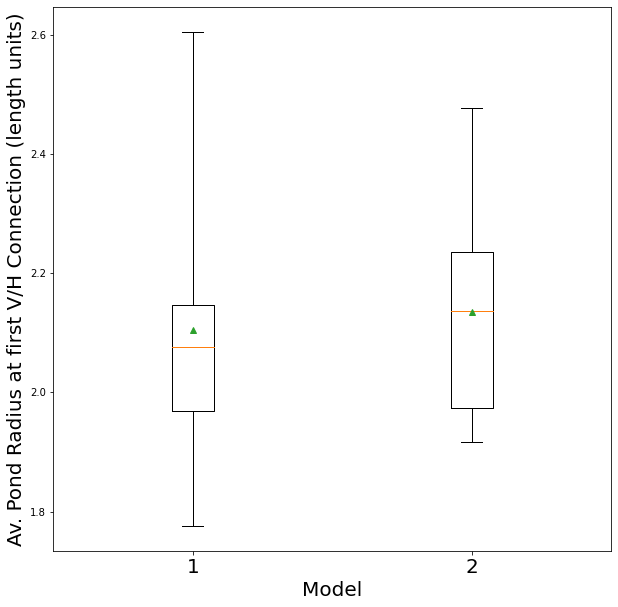}}
  
  \end{center}
  \caption{\centering Plots for $r_{perc}$}
  \label{fig:percolation}
\end{figure*}

Next, we consider the results from the calculation of $r_{perc}$, the minimum radius out of channels which traverse the sample region from one edge of the image to the opposite edge. $r_{perc}$ is plotted for each model separately in Fig.~\ref{fig:percolation}. The time step at which this value becomes positive marks the time at which both pairs of opposite sides become connected by ponds. This can be interpreted as the time when full percolation begins for the pond system. We see that for Model 3 this threshold is only reached by a single run, which suggests that percolation is happening on smaller length scales, and that either drainage events happen too soon or too easily to generate full percolation across the image.

Combining the information about when percolation begins with the other information obtained from the signed Euclidean distance filtration, we can plot the percolation threshold $r_{perc}$ versus the pond coverage fraction and average pond radius, as well as plotting it over time. These are plotted in Figs.~\ref{fig:percolation-d},\ref{fig:percolation-e} and \ref{fig:percolation-f}.

For Models 1 and 2 we plot a box-and-whiskers graph of the times at which percolation begins (Fig.~\ref{fig:percolation-g}). We can see that while the mean hitting time for both models is similar, Model 2 shows slightly less variability and a later overall hitting time. This is consistent with other results, as Fig~\ref{fig:spanning-a}-\ref{fig:spanning-b} also point to greater overall percolation in Model 1. This may suggest that ice-albedo feedback results in percolation occurring later than if the feedback was not present, or it may be an artifact of the dimension normalisation: at a fixed total melt rate, isolated ice between ponds will melt slower in the feedback model.  Model 3 is excluded because it does not achieve cross-image percolation.

We also analyse pond coverage fraction and average pond radius when percolation begins (Figs.~\ref{fig:percolation-h},\ref{fig:percolation-i}), which provides the time step at which global connectivity occurs as well as critical area/pond radius threshold for percolation. Our results in the context of other percolation results for ponds offers a critical area fraction estimate $\phi_{c}\approx41\%$ for the first model and $\phi_{c}\approx43\%$ for the second, while \citet{FractalDimensionPonds} finds a critical area fraction $\phi_{c}\approx25\%$. We remark  \citet{FractalDimensionPonds} computes this critical area fraction using fractal dimension rather than with connectivity of the sample region. 
Besides differences in the measurement methods, the difference in critical area fraction suggests that compared with real data, our model generates ponds that are substantially more compact or disconnected at the same area, which would be expected from a spherical pond growth model. Since the critical threshold in pond coverage for Model 3 is between the $\phi_c$ estimates for the real data and Models 1 and 2, this could indicate that a more realistic pond growth equation would likely produce full-length percolation in a drainage model, replicating real results.


As with our first method for detecting spanning channels, we see also a particularly strong signal of a critical pond radius in Fig.~\ref{fig:percolation-f}, at roughly $2.1$ length units (Fig.~\ref{fig:percolation-f}). We note that the model is dimensionless. Seeing the same critical threshold in both methods suggests that the length scale of percolation may be less important than local connectivity for these measurements.

Robust investigation of real data, or a wider variety of models, developing an appropriate normalisation and statistical testing, would be necessary to assert any of these claims with certainty. Our work is preliminary and exploratory, and we leave these questions for future work. 


% ...

\section{\label{sec:discussion}Conclusion and future Work} 
We showed that topological data analysis applied to melt pond images yields detailed information about connectivity and geometry of pond networks, which goes beyond existing summaries like coverage fraction and fractal dimension. We are able to determine the minimum radii of channels that span a given image, doing so within a broader framework that gives information about many aspects of pond structure, such as the number and aspect-ratio of concave ponds. We work toward a model selection pipeline using these detailed summary statistics.


Although we tested many possible topological summaries for this type of data, and showed clear functional differences corresponding to the topological invariants presented, there are still several promising directions for investigation. 

One fact to bear in mind about our method for determining node and throat radii for concave pond structures is that a single connected pond component can yield multiple birth/death pairs corresponding to the different node-like subponds contained within a pond. While we believe this is a benefit for investigating the network-like structure of melt ponds, it would also be interesting to partition the set of birth/death pairs in regions A and B of the 0\textsuperscript{th} homology persistence diagram into those features which arise within the same pond component. Doing so would allow us to find the minimum and maximum radius of a given pond component, and compare the two. We believe this can be done by tracking which 0-cycles merge with which others and grouping them accordingly, although we have not implemented this.

Another piece of information obtained for binary images in \citet{RobinsSEDT} is the distance between where birth and death events happen in the image for each birth/death pair. They use this as another way to detect percolation, which could be scale invariant. In addition, in that paper the authors work within the framework of discrete Morse complexes, allowing them to determine the difference to the phase boundary itself, rather than the nearest pixel of the opposite phase - meaning their results are less affected by differences in image resolution. In the supplementary material, we discuss the effect of image resolution in more detail.

Often, real-world data is noisy, and we would like some guarantee that the statistical outputs of our methods are stable with respect to certain forms of noise. Section D of the supplementary material discusses stability results for the signed Euclidean distance transform with respect to certain relevant types of noise, but there are still forms of noise for which our method is not stable. Notably so-called `salt and pepper' noise can drastically change the signed Euclidean distance of a binary image: introducing a pixel of one phase into a large region of the opposite phase can wildly affect the SEDT value of nearby pixels. Salt and pepper noise is a well-known problem in TDA, and some methods, like \textit{multiparameter persistent homology} (MPH) have been developed to deal with it, although currently these methods are computationally intensive. In MPH, we filter a simplicial complex by a second parameter, often density, in addition to our original filtration, resulting in a \textit{bifiltration}. Perhaps incorporating MPH into our framework via an SEDT/density bifiltration could improve the robustness of our method to noise.

While we applied our methods to time-series data, the topological data analysis methods we used are not themselves time-dependent. Thanks to recent computational advances \cite{zigzag}, it is now possible to track how topological features in a time-series change over time using \textit{zigzag homology}. We believe that this method could also be useful for studying continuous melt pond time series like those generated by our model, giving insight into the distribution of melt pond dynamics in both space and time, although in real-world data it is unlikely that we would have a time-series with frequent enough measurements to apply zigzag homology.

Previously, percolation behaviour for melt ponds has been investigated via fractal dimension. Interestingly, fractal dimension can be inferred from the growth of topological features in differently-sized subsamples \cite{fractaldimdefinition}, giving an alternative estimation of fractal dimension in data. In section A of the supplementary material, we summarize this definition of fractal dimension and apply the methods in \citet{JAQUETTE2020105163} to compute this for pond data.

\section*{Acknowledgements}
WO thanks EPSRC Vacation Internship Fund, St. John's College James Fund and the Mathematical Institute at Oxford for providing a stimulating environment for this undergraduate research experience. HAH gratefully acknowledges funding from a Royal Society University Research Fellowship.
GG and HAH are members of the Centre for Topological Data Analysis, which is funded by the EPSRC grant `New Approaches to Data Science: Application Driven Topological Data Analysis' \href{https://gow.epsrc.ukri.org/NGBOViewGrant.aspx?GrantRef=EP/R018472/1}{\texttt{EP/R018472/1}}. GG acknowledges funding as well from NSF MSPRF grant 2202895.



\bibliography{Bibliography}
\section*{Supplementary Material}

\section*{A: Fractal Dimension Via Persistent Homology}\label{appendix fractal}
\begin{figure}[h]
 \begin{center}
 \subfigure[Rips Filtration]{\label{fig:rips-a}\vbox{\offinterlineskip\halign{#\hskip3pt&#\cr
  \includegraphics[width=0.2\columnwidth]{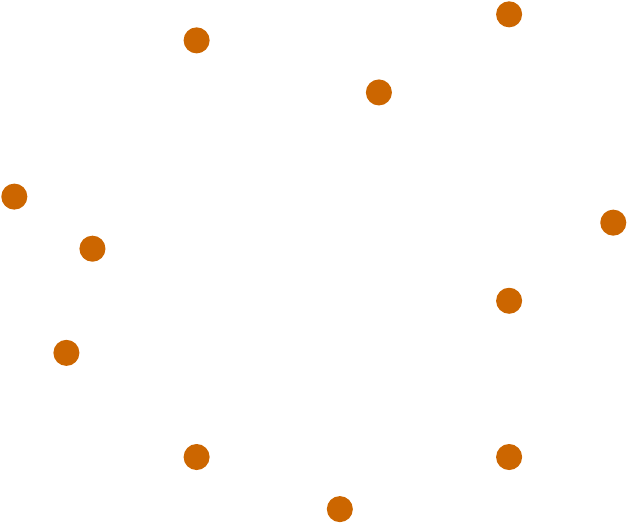}&
  \includegraphics[width=0.2\columnwidth]{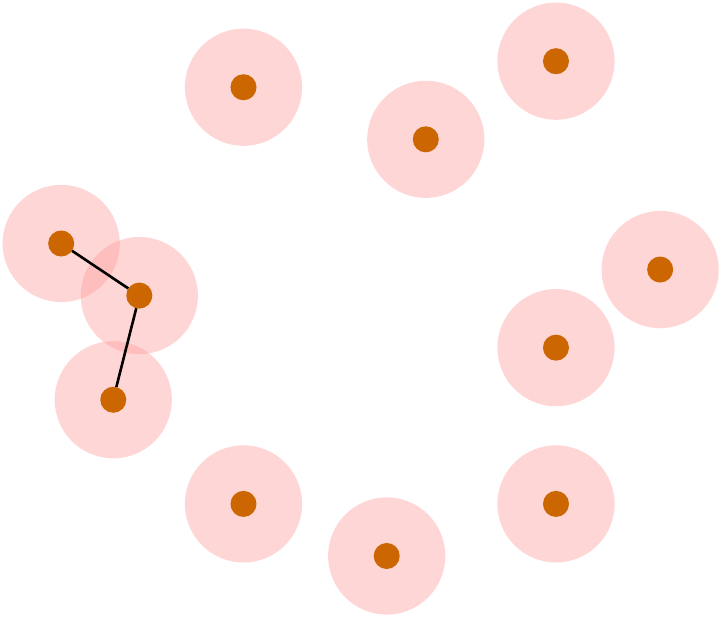}\cr
  \noalign{\vskip3pt}
  \includegraphics[width=0.2\columnwidth]{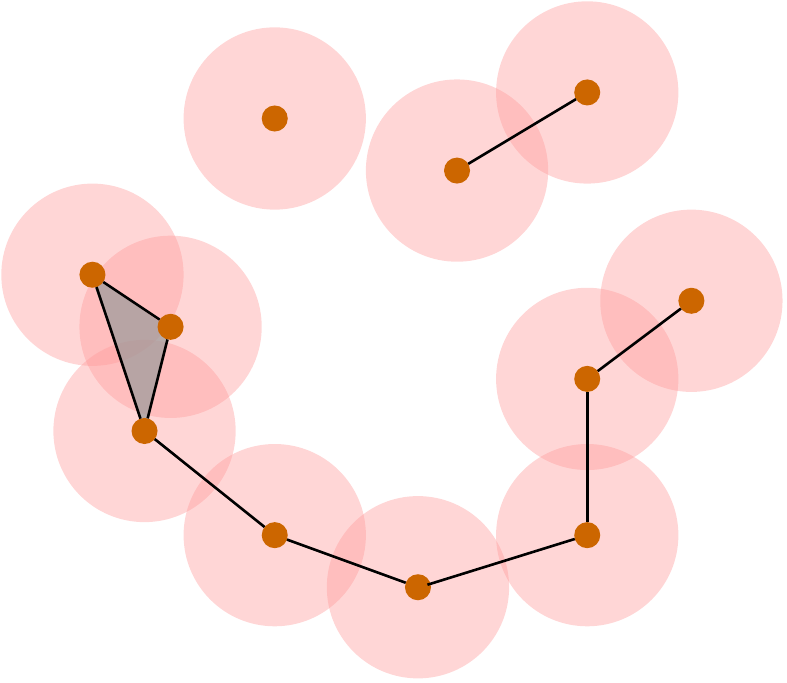}&
  \includegraphics[width=0.2\columnwidth]{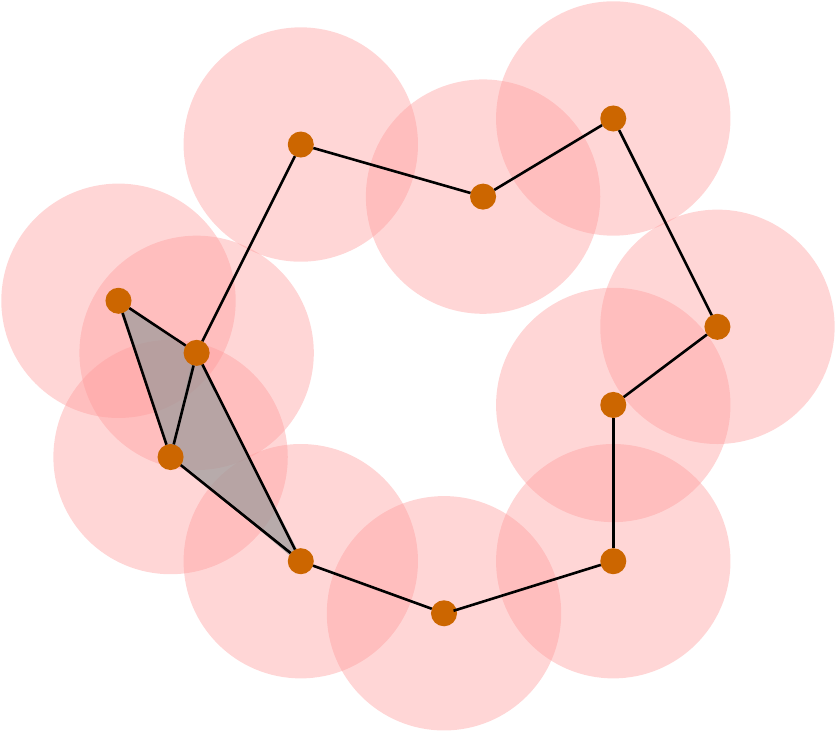}\cr
  }}}
  \subfigure[Persistence Diagram]{\label{fig:rips-b}\includegraphics[width=0.4\columnwidth]{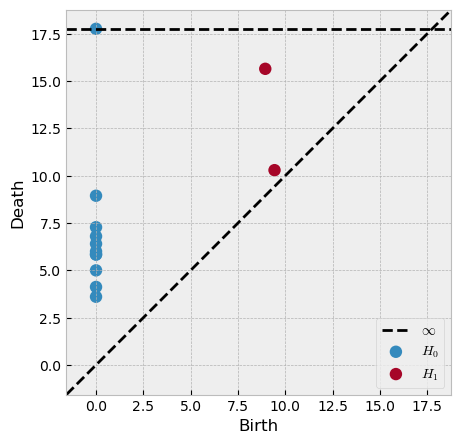}}
  \end{center}
  \caption{\centering Rips Filtration: A simplicial complex is built up by connecting points of pairwise distance below a threshold}
  \label{fig:rips}
 \end{figure}

\textit{Persistent Homology Fractal Dimension} is a family of definitions of fractal dimension for measures introduced in \citet{fractaldimdefinition}. Unlike other definitions of fractal dimension, Persistent Homology Dimension is based on the limiting behaviour of the persistence diagrams for a filtration based on subsamples taken from the measure $\mu$. This filtration is known as the \textit{Vietoris-Rips} filtration and is widely used in TDA. Starting from a point cloud sampled from $\mu$, we grow balls of radius $\varepsilon$ centred on each point, connecting points pairwise when these balls intersect, and adding $n$-simplices wherever a subset of $n+1$ points are completely connected. In this way, we form a simplicial complex for each value of $\varepsilon$, resulting in a filtration to which we apply persistent homology. (Fig.\ref{fig:rips}).

\noindent Given a subset X of a metric space, for each dimension i$\geq$0 we define $$E^{i}_{\alpha}(X):=\sum_{(b,d)\in PH_{i}}|d-b|^{\alpha}$$ where $PH_{i}$ is the persistence diagram for the i-th homology and $\alpha$ is some constant $>$0. Letting X=$\{x_{1},x_{2},...,x_{n}\}$, a random sample from $\mu$, we define the i\textsuperscript{th} persistent homology dimension as: $$dim_{PH_{i}^{\alpha}}:=\frac{\alpha}{1-\beta}$$ where $$\beta:=\limsup_{n\rightarrow\infty}\frac{\log(\mathbb{E}(E_{i}^{\alpha}(\{x_{1},x_{2},...,x_{n}\})))}{\log(n)}.$$

This ensures that $E_{i}^{\alpha}$ scales like $n^{\frac{d-\alpha}{d}}$ where $dim_{PH_{i}^{\alpha}}=d$. In practice, we estimate $dim_{PH_{i}^{\alpha}}$ via computing $E_{i}^{\alpha}$ for a range of sample sizes, and fitting a regression line to the graph of $\log(E_{i}^{\alpha})$ vs $\log(n)$. \citet{JAQUETTE2020105163} finds that $dim_{PH_{0}}$ converges much faster than $dim_{PH_{1}}$, and converges fastest for smaller values of $\alpha$. For a measure defined on $\mathbb{R}^{2}$, $dim_{PH_{i}}=0$ for $i\geq2$.

\citet{fractaldimdefinition} and \citet{JAQUETTE2020105163} show that in computational experiments, persistent homology fractal dimension approximately agrees with other definitions of fractal dimension, like Hausdorff, correlation and box-counting dimensions, for certain cases like the Sierpinski triangle. \citet{fractaldimdefinition} also shows that for a measure $\mu$ on $X\subseteq\mathbb{R}^{m}$ for $m\geq2$, $dim_{PH_{0}}\geq m$, with equality if the absolutely continuous part of $\mu$ has positive mass, and makes a number of conjectures about the theoretical properties of $dim_{PH_{i}}$. \citet{JAQUETTE2020105163} computes PH dimension for a range of self-similar fractals, strange attractors in chaotic systems, and real-world earthquake data. 

We compute persistent homology fractal dimension for both the melt pond interiors and boundaries. We take the uniform measure on the subset of each binary image corresponding to ponds. We randomly sample points from the set of pond pixels. Importantly, if the pixel size is 1 unit we add uniform random noise in the interval $(-0.5,0.5)$ to ensure that points are sampled uniformly from the entire submanifold of $\mathbb{R}^{2}$ spanned by the ponds. Not doing so results in anomalies due to image resolution.

\begin{figure*}

\subfigure[Regression plot for $dim_{PH_{0}}$ for one timestep]{\label{fig:fractaldim-a}
\includegraphics[width=0.4\textwidth]{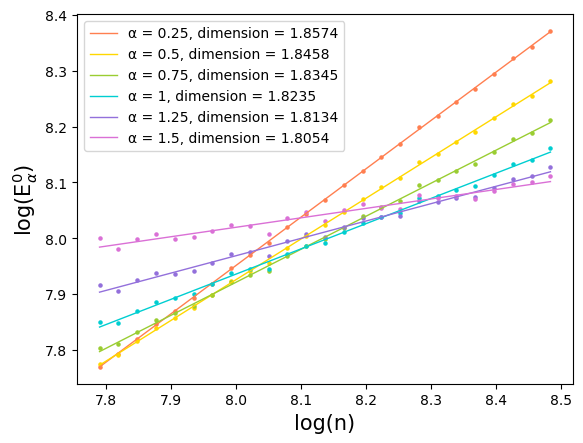}}
\subfigure[Regression plot for $dim_{PH_{1}}$ for one timestep]{\label{fig:fractaldim-b}
\includegraphics[width=0.4\textwidth]{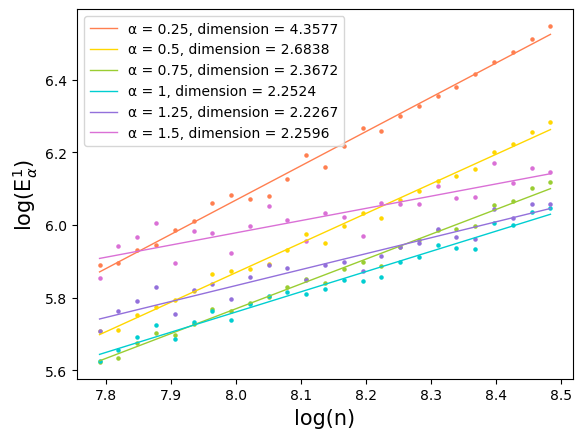}}
\subfigure[Plot of average $dim_{PH_{0}}$ of pond interior against time]{\label{fig:fractaldim-c}
\includegraphics[width=0.4\textwidth]{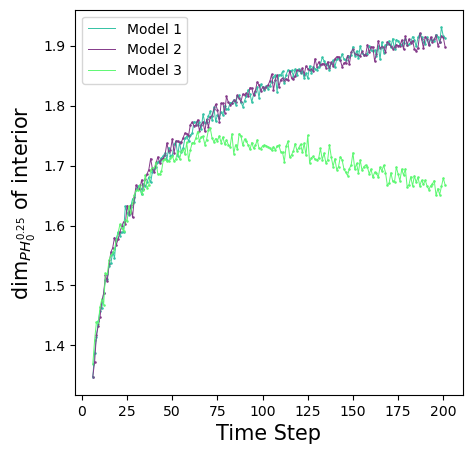}}
\subfigure[Plot of average $dim_{PH_{0}}$ of pond boundary against time]{\label{fig:fractaldim-d}
\includegraphics[width=0.4\textwidth]{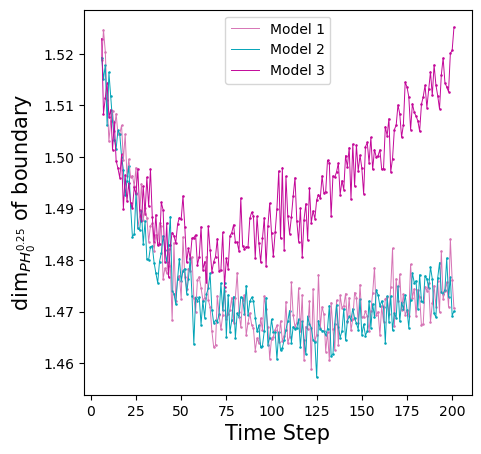}}
\caption{Persistent homology fractal dimension plots for model runs}
\label{fig:fractaldim}
\end{figure*}

We plot the results of persistent homology fractal dimension for the model data in Figure~\ref{fig:fractaldim}. \ref{fig:fractaldim-a},\ref{fig:fractaldim-b} show that while 0\textsuperscript{th} homology fractal dimension appears to converge for small values of $\alpha$, 1\textsuperscript{st} homology fractal dimension is not close to converging for our chosen range of sample sizes. Average fractal dimension of pond interiors and boundaries is plotted as a function of time and pond coverage area in \ref{fig:fractaldim-c},\ref{fig:fractaldim-d}. We do not expect to see the same transition as observed in \citet{FractalDimensionPonds}, since in this paper fractal dimension was calculated for individual ponds, whereas we calculate it over an entire image. What we see is an initial decrease in fractal behaviour (for the interior, a fractal dimension closer to 2 is less fractal, while for the boundary, a fractal dimension closer to 1 is less fractal). This probably indicates that the multiple connected components in the early stages of pond evolution cause more fractal behaviour to be registered by persistent homology fractal dimension. In other words, the approach of calculating fractal dimension on the entire image rather than individual ponds does not give fine enough detail to observe any percolation threshold. To investigate further a critical area theshold for persistent homology fractal dimension, we should follow \cite{FractalDimensionPonds} and calculate the fractal dimension of individual ponds.

\section*{B: Computing Spanning Channels}\label{appendix spanning}
We would like to compute the number and throat radius of channels which span the image from edge to edge, rather than those which form `internal' loops in the interior of the image. To do this, we can compute the death values for birth/death pairs in region A of the SEDT superlevel filtration (the channel radii between separate connected components in the ice phase), and remove from this list the birth values for birth/death pairs in region C of the SEDT sublevel filtration (the channel radii of loops in the pond phase). However, two situations can cause the problem to arise that no death value in the superlevel 0\textsuperscript{th} homology corresponds to a given birth value in the SEDT sublevel 1\textsuperscript{st} homology. We deal with these below.

\begin{figure}[h]
\subcapcentertrue
  \begin{center}
  \subfigure[Issue with diagonally connecting ice]{\label{fig:problems-a}\includegraphics[width=0.3\columnwidth]{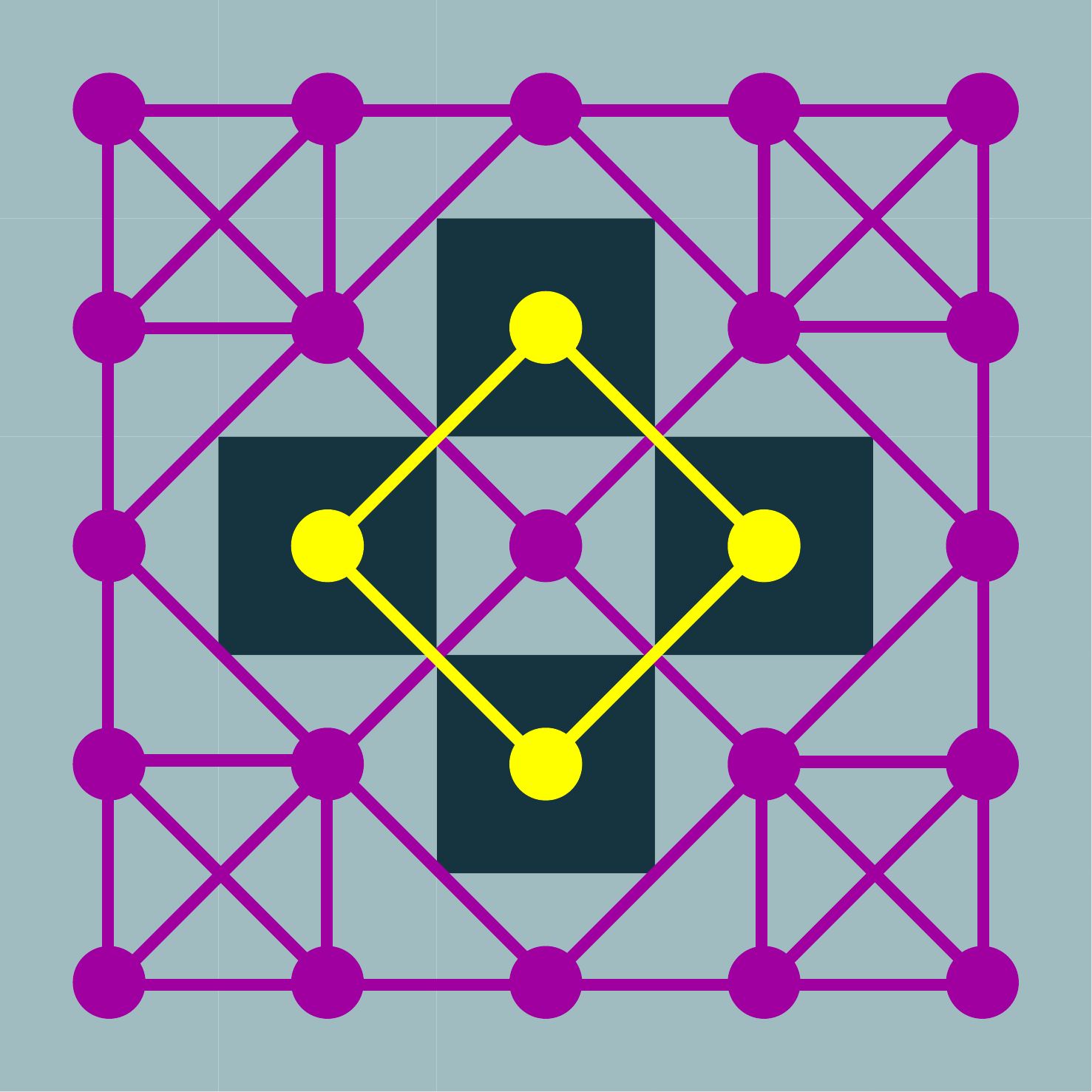}}
  \subfigure[Issue with boundary conditions]{\label{fig:problems-b}\includegraphics[width=0.3\columnwidth]{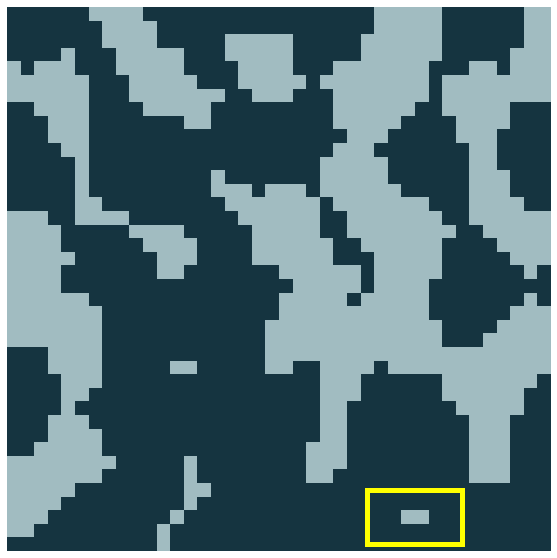}}        
  \end{center}
  \caption{\centering Issues which cause birth values for loops in pond structure to have no corresponding death value of ice 0-cycles}
  \label{fig:problems}
\end{figure}

First, consider the arrangement in Fig.\ref{fig:problems-a}. By default, in both the superlevel and sublevel filtrations, we connect each pixel to its 8 neighbours, including diagonal neighbours. However, in the situation depicted, we can see that while the 4 pond pixels are connected into a loop, the interior ice pixel is simultaneously connected to the ice outside the loop, resulting in a pond loop with no connected component of the ice phase enclosed inside it.

To fix this, we make the arbitrary choice to 8-connect the sublevel filtration while 4-connecting the superlevel filtration. This ensures that in cases like those depicted, the interior ice pixel is cut off from the rest of the ice, as desired.

Secondly, suppose we have a large region of water near the boundary of an image, with an island of ice near the boundary so that its distance to the image boundary is smaller than its distance to the nearest distinct ice component (Fig.~\ref{fig:problems-b}). In this situation, note that there will be a loop in the pond phase who's birth value is the distance from the ice island to the boundary. However, at this value of the superlevel filtration, there is no death event corresponding to when the 0-cycle hits the boundary.

We can deal with this by introducing some boundary conditions which ensure that when a 0-cycle hits the boundary it dies. We want a 0-cycle to die in any of the following cases:
\begin{enumerate}
    \item an ice component which is cut off from the image boundary should die when it meets another ice component, or the image boundary.
    \item an ice component which is adjacent to the image boundary should die when it meets another ice component which is adjacent to the image boundary.
\end{enumerate}

\begin{figure}[htp]
  \begin{center}
    \includegraphics[width=0.5\columnwidth]{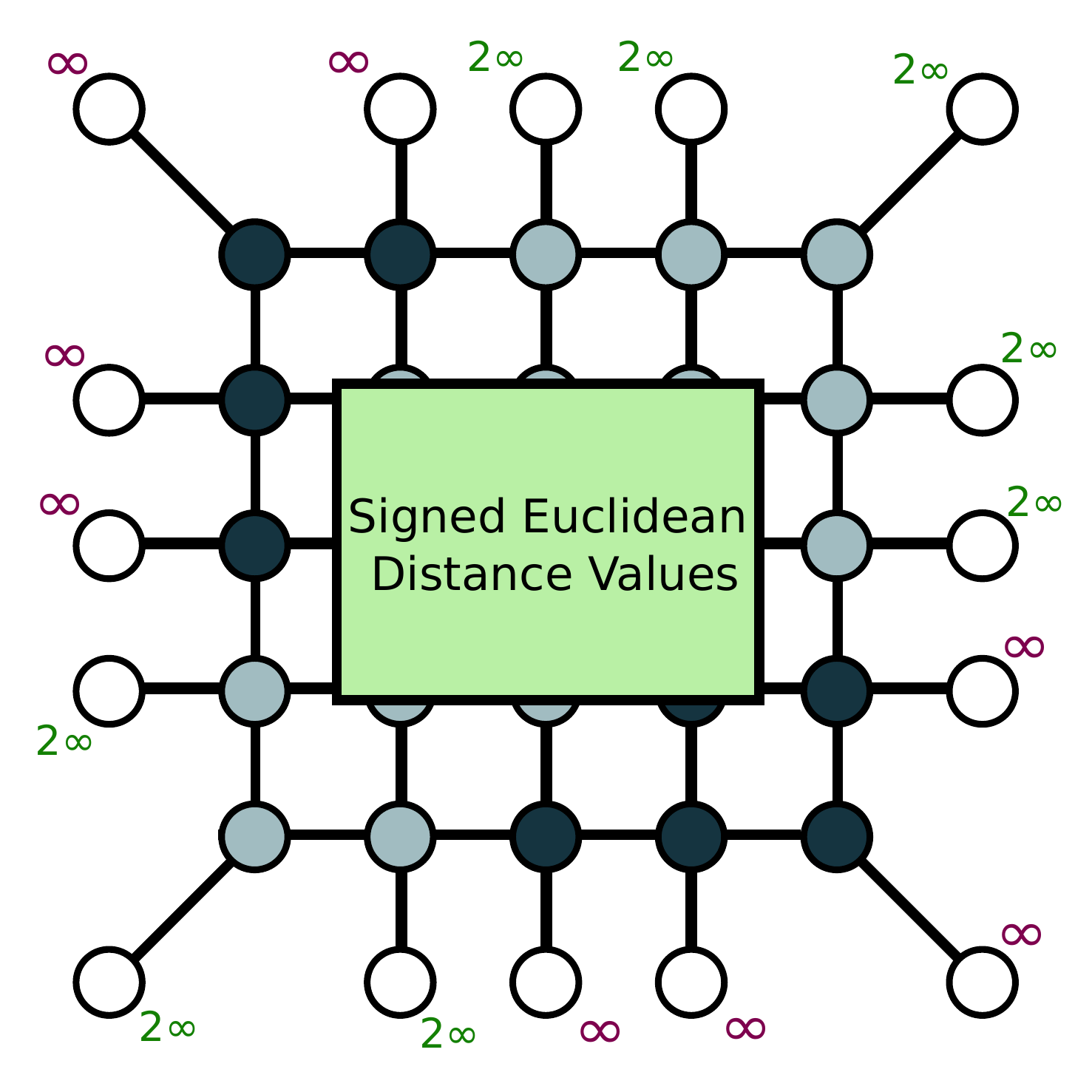}
    
  \end{center}
  \caption{\centering Boundary conditions for the superlevel SEDT filtration. We append boundary vertices to each pixel on the image border. If the vertex is connected to a water pixel, we give it a filtration value of $\infty$. If the vertex is connected to an ice pixel, we give it a filtration value of $2\infty$.}
  \label{fig:boundaryconditions}
\end{figure}

To achieve this, we take the filtered simplicial complex arising from the signed Euclidean distance transform and add a new vertex connected to each pixel on the image boundary. If the boundary pixel the vertex is connected to is in the pond phase, we give this new vertex a filtration value of $\infty$. If the boundary pixel is in the ice phase, we abuse notation and write the filtration value as $2\infty$ to indicate that these vertices should be be born before all image pixels and before the vertices of filtration value $\infty$. (Fig.\ref{fig:boundaryconditions}) These boundary conditions guarantee that the birth values for the 1\textsuperscript{st} homology of the pond phase are a subset of the death values for the 0\textsuperscript{th} homology of the ice phase, as required. We are interested in just those birth/death pairs in the difference of the two sets with birth value either finite or $2\infty$. Features with birth value $\infty$ correspond to the additional boundary vertices we added that connect to pond pixels, and we ignore these.

\section*{C: Computing $r_{perc}$}
Given a binary image, we can also compute $r_{perc}$, the minimum throat radius of channels which span from one side of the image to the opposite side. 

To do this, we break the task into detecting the minimum radius $r_{ver}$ of channels spanning from the top edge to the bottom edge, and the minimum radius $r_{ver}$ of channels spanning from left to right. To compute $r_{ver}$ we can add a 1-pixel boundary to the top and bottom edges with filtration value $-\infty$. These boundary components will always be present in the sublevel filtration, but will merge into a single component at some filtration value $d$. Given the persistence diagram for this filtration, then, there will be a unique birth-death pair with birth value $-\infty$ and finite death value $d$. If the death value $d$ is negative, we know that there is a pond channel which spans the image vertically, and the minimum throat radius of this channel is $-d$. We define $r_{ver}:=-d$. If no channel spans the image from top to bottom edge, then the death value will be positive. This death value then measures the largest radius of ice that a test particle would need to traverse to cross the image from one side to another.

Similarly, we can add boundaries to the left and right edges of the image with filtration value $-\infty$. We can then determine the death value of the component corresponding to these boundaries, and call this value $r_{hor}$ - this is the minimum throat radius of channels which span horizontally from the left to the right edges of the image. We define $r_{perc} := \max\left\{r_{ver},r_{hor}\right\}$, which then tells us the radius of the smallest sphere which can traverse the image from one side to the opposite either vertically or horizontally. Again, if $r_{perc}$ is negative, it measures the largest radius of ice which a test particle must cross to traverse the image. The first time step at which $r_{perc}>0$ tells us the time at which either pair of opposite edges is first connected, that is, this marks when percolation begins.

\section*{D: Stability Results for the Signed Euclidean Distance Transform}\label{appendix stability}


Of central importance in the theory of persistent homology is stability: real-world data is often noisy, and so we would like some guarantees of the types of transformation under which our topological statistics change only by a small perturbation. A basic stability result guarantees that persistence diagrams are stable under small changes of the underlying filtration function (with respect to the so-called `bottleneck distance' metric on persistence diagrams - see \cite{Cohen-Steiner2007} for a definition). 
Therefore, we would like some idea of the transformations we can make to a binary image under which the change in the signed Euclidean distance is bounded. 

\citet{Resolution} proves that for signed Euclidean distance functions $g_{1}(x)$ and $g_{2}(x)$ defined for two subsets  $X_{1}$ and $X_{2} \subseteq \mathbb{R}^{n}$, the $L^{\infty}$ norm $||g_{1}-g_{2}||$ between them is bounded above by a bound:

$$B=\max\left\{\begin{aligned}sup_{x_{2}\in X_{2}^{c}}d(x_{2},X_{1}^{c})+sup_{x_{1}\in X_{1}}d(x_{1},X_{2}),\\
sup_{x_{1}\in X_{1}^{c}}d(x_{1},X_{2}^{c})+sup_{x_{2}\in X_{2}}d(x_{2},X_{1})\end{aligned}\right\}$$

\noindent which is less than the sum of the Hausdorff distances $\mathcal{H}(X_{1},X_{2})+\mathcal{H}(X_{1}^{c},X_{2}^{c})$ and in the case where one of the sets has boundary with positive reach $r>\max\left\{\mathcal{H}(X_{1},X_{2}),\mathcal{H}(X_{1}^{c},X_{2}^{c})\right\}$, this bound can be tightened to the maximum of the Hausdorff distances. 

This bound on the $L^{\infty}$ norm of the signed Euclidean distance functions guarantees a bound on the bottleneck distance between the resulting persistence diagrams:

$$\text{\textbf{Bottleneck}}(Dgm(g_{1}),Dgm(g_{2}))\leq B$$

While this result holds in the continuous setting, for signed Euclidean distance functions defined on all of $\mathbb{R}^{n}$ and with respect to subsets of arbitrarily fine detail, \citet{Resolution} discusses the discrete setting of image data, and finds that provided the image resolution is high enough, we can bound the bottleneck distance between the persistence diagrams of continuous and discrete SEDT filtrations. This gives some reassurance that our discrete method is a good approximation to the underlying ground truth of the melt ponds.

In practice, what the stability result above does is guarantee that changing the values of a binary image within a small neighbourhood of the phase interface will only affect the resulting persistence diagram a small amount. This is useful in the case of melt ponds, since often there is some slushy snow around the boundary of the ponds, which can be mislabelled as pond water by image processing techniques.

Although we have stability with respect to the bottleneck metric on persistence diagrams, the statistics we extract from persistence diagrams may still be unstable. The bottleneck metric gives low weight to $(\text{birth},\text{death})$ pairs which are close to the diagonal, and hence persistence diagrams which vary greatly in statistics like number of points and average value of birth/death can nevertheless have small bottleneck distance between them. To improve the stability of our statistics, we could filter points by persistence, or weight birth/death values by persistence.

\end{document}